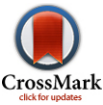

PLOS ONE

# Method for Finding Metabolic Properties Based on the General Growth Law. Liver Examples. A General Framework for Biological Modeling

## Yuri K. Shestopaloff*

Segmentsoft Inc., Toronto, Ontario, Canada

## Abstract

We propose a method for finding metabolic parameters of cells, organs and whole organisms, which is based on the earlier discovered general growth law. Based on the obtained results and analysis of available biological models, we propose a general framework for modeling biological phenomena and discuss how it can be used in Virtual Liver Network project. The foundational idea of the study is that growth of cells, organs, systems and whole organisms, besides biomolecular machinery, is influenced by biophysical mechanisms acting at different scale levels. In particular, the general growth law uniquely defines distribution of nutritional resources between maintenance needs and biomass synthesis at each phase of growth and at each scale level. We exemplify the approach considering metabolic properties of growing human and dog livers and liver transplants. A procedure for verification of obtained results has been introduced too. We found that two examined dogs have high metabolic rates consuming about 0.62 and 1 gram of nutrients per cubic centimeter of liver per day, and verified this using the proposed verification procedure. We also evaluated consumption rate of nutrients in human livers, determining it to be about 0.088 gram of nutrients per cubic centimeter of liver per day for males, and about 0.098 for females. This noticeable difference can be explained by evolutionary development, which required females to have greater liver processing capacity to support pregnancy. We also found how much nutrients go to biomass synthesis and maintenance at each phase of liver and liver transplant growth. Obtained results demonstrate that the proposed approach can be used for finding metabolic characteristics of cells, organs, and whole organisms, which can further serve as important inputs and constraints for many applications in biology (such as protein expression), biotechnology (synthesis of substances), and medicine.

## Introduction

The general growth law is one of the major mechanisms of Nature governing evolutionary development of living species and their individual growth and development at different spatial levels, from cellular components to whole organisms [1]. It is also that mechanism, which is responsible for balanced growth of organs and systems within the same multicellular organism. In the previous studies, a method based on the general growth law has been proposed for modeling growth of organs and finding different growth parameters, such as change of mass, size, geometrical characteristics, distribution of replicating and quiescent cells (exemplified by hepatocytes in livers), etc. The practical value of the proposed method was demonstrated by considering growth characteristics of livers and liver transplants.

Here, we present a continuation of this study, introducing a general method for finding metabolic parameters, such as nutrient influx consumed by growing and grown cells and organs in order to support different functions. (Nutrient influx is the rate at which nutrients are consumed by organisms or its constituents.) Once we know nutrient influx, we can find integral metabolic characteristics, such as accumulated amount of nutrients required to support biomass synthesis and maintenance needs for each phase of growth or for a certain period. We demonstrate method's efficiency by applying it to study of liver metabolism, using experimental data on growth of liver transplants in dogs [2], when whole small livers were transplanted from small dogs to big dogs, and liver grafts in recipients and liver remnants in donors in case of humans [3,4]. Then, using results of our studies, as well other published materials, we introduce a general framework for modeling biological phenomena in general, and discuss its possible application to a recently launched grand undertaking on creating a comprehensive biochemical and biophysical liver model - the Virtual Liver Network project [5]. In this, we will aim at obtaining robust, transparent, comprehensive and practical framework for modeling biological phenomena, composed of interrelated methods, models and conceptual approaches. This general framework introduces systemic approach that would allow solving a wide range of practical and theoretical problems in biology, biotechnology, medicine and other disciplines.

## 1. Liver Metabolism

The definition of metabolism in Webster's dictionary as "the sum of the processes by which particular substance is handled in the living body" is close to intuitive understanding of metabolism phenomena by most scientists. However, the same term is often used when one refers to entirety of biochemical reactions in a living organism or its constituent, like cell metabolism, while specifying a particular metabolic mechanism when needed, such as energy metabolism. In our case, we consider the overall liver metabolism from the perspective of nutrient consumption for biomass synthesis, on one hand, and maintenance of existing biomass on the other.

The role of liver metabolism in mammals is extremely important. All nutrients absorbed from gastrointestinal tract pass through the liver, which processes nutrient components and stores some substances. Liver is one of the major organs supporting organism homeostasis. For instance, it produces and maintains level of glucose in organism through certain signaling feedbacks. Metabolic failure of liver leads to animal death. Liver is responsible for many metabolic functions, such as oxidizing triglycerides to produce energy, synthesize lipoproteins, cholesterol, phospholipids, converting excessive carbohydrates and proteins into fatty acids and triglyceride, maintaining stable level of glucose, removing ammonia through urea synthesis, breakdown of toxic substances, as well as supporting many other functions [6].

Liver metabolic studies relate to the following areas: diseases of liver and other organs, since liver metabolic disorders affect other organs and vice versa; medical studies, including liver transplantation issues and therapeutic curing; drug metabolism for medical and pharmacological applications; numerous biological studies. For these purposes, different models of liver metabolism, on cellular and above levels, have been developed. For instance, work [7] models iron metabolism in livers. The model is based on analysis of a hepatic biochemical network involved in iron sensing and regulation. The authors used in vitro biochemical data, such as protein complex dissociation constants. Mathematically, the model is described by system of ordinary differential equations which include the rate of change of chemical substances. In [8], authors introduce a liver model on the basis of reaction kinetics, solving the system of differential equations describing dynamic mass balance, thus finding the concentration dynamics of different metabolites and intermediate products (mostly related to gluco-neogenesis) in the blood and tissue. Similar approaches are used in liver cell bioreactors for liver support therapies. Kinetic models are useful for practical applications, such as predicting occurrence of hypoglycemic events in rigorous insulin therapy [9]. Another example could be using models for human liver cell lines for xenobiotic metabolism and toxicity studies [10].

These and many other similar studies consider metabolic pathways, sometimes united on the basis of groups of related metabolites. However, all involved biochemical reactions do not exist independently but are tightly interconnected. The whole liver biochemistry is a single machine, regardless of whether it serves growth, liver's own maintenance requirements or needs of other organs and systems. In such an arrangement, it is important to know distribution of nutrient resources between different functions, such as how much nutrients go to biomass synthesis, or used for liver's own maintenance needs, what is the amount of nutrients that go to the rest of an organism, and so on.

This goal of the study is illustrated by Figure 1 that shows distribution of nutrient influxes. Eventually, if one wants to have a complete picture of the liver metabolism, all metabolic pathways have to be linked to these integral parameters. Unfortunately, it is very difficult to find out how much nutrients are consumed by the liver itself, how much nutrients are used for maintenance needs and for biomass production. In other words, nutrient influxes 1–4 shown in Figure 1 can not be quantified. Besides, an important issue is defining interdependencies between biochemical reactions and appropriate nutrient influxes that serve growth and liver maintenance needs (a box "Liver metabolic pathways" in Figure 1). The problem is that all nutrients go through the liver, both for the liver consumption and for the rest of an organism. Application of the general growth law allows solving this important problem.

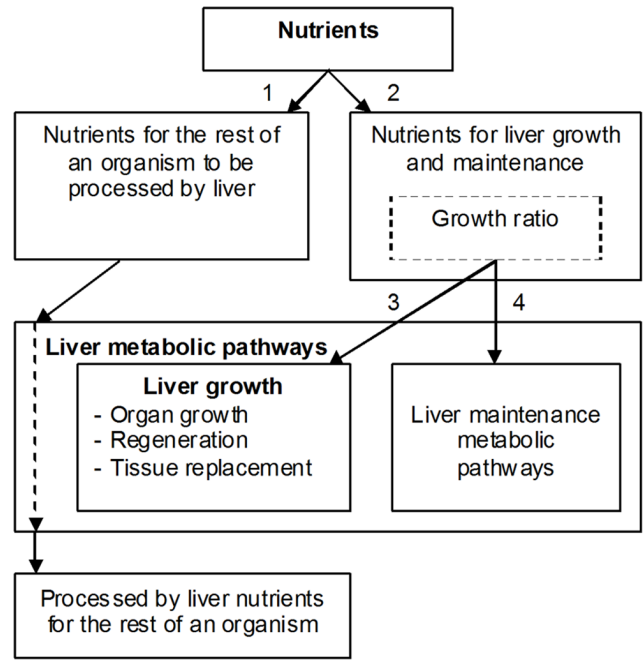

**Figure 1. Distribution of nutrients in liver.** Nutrients are used for biomass synthesis and maintenance needs in the liver. Processed nutrients go to the rest of an organism.
doi:10.1371/journal.pone.0099836.g001

## 2. Specifics of Biological Modeling

An example of presently used approaches can be work [11], in which the authors introduce a model of a growing liver. It incorporates several biomolecular mechanisms affecting liver cells, which are divided into three groups: quiescent (Q), primed (P) - cells that have switched to G1 phase, but do not necessarily proceed to replication, and replicating (R) cells. The authors describe their approach as follows: "The rates of change of cell number are modeled in terms of signaling molecules interacting with cells in each state Q, P, and R. The simplest assumption for these reactions is that they proceed following the law of mass action. As such, the transition rate between states is proportional to the number of signaling molecules and to the fraction of cells in the state affected by the signal. This leads to second-order steps in the transition equations." The authors further note that "The signals themselves have nonlinear rates of increase and decrease governed by Michaelis-Menten dynamics." The model provides realistic predictions and, as authors assert, complies with experimental data, although no direct comparison with particular experimental observations has been shown. There are some counter-intuitive inferences from the obtained results. For instance, two hepatectomy resections, according to that model, provide faster liver regeneration than one hepatectomy, which is doubtful given the doubled recovery period. The model is an interesting development, although its practical value remains questionable, because of the significant amount of required input data, many of which are not readily available, and the set of assumptions, whose validity yet remains to be proven.

Numerous publications of collaborators at Virtual Liver Network (VLN) project related to growth of cells and tissues can be found on the project's website (http://www.virtual-liver.de/). The appealing feature of the project is its claimed practical orientation through application of methods of system biology and system medicine "to demonstrate an impact on the needs of patients and clinical practitioners alike". In the context of our study, the project's emphasis on "prototyping ways to achieve true *multi-scale* modeling within a single organ and linking this to human physiology" is very meaningful, demonstrating a healthy trend towards systemic approach, which is especially important for living organisms, in which workings of all mechanisms and systems closely interrelate. However, the strategies for implementing this general idea are yet to be defined. In this regard, the general growth law, as a *fundamental* law of Nature, acts at different scale levels (in the same way as Newton's laws of mechanics are valid in a wide range of dimensional sizes - from a micro-world to cosmic bodies). Knowing such fundamental laws of Nature is crucial for defining strategic directions and conducting efficient studies in biology. The reason is the tight interconnection of different biochemical and biophysical mechanisms, acting at different scale levels. Besides, these mechanisms are supported by numerous interrelated regulative and reactive feedback loops acting within the same organism at different scale levels. Such an arrangement of life creating and maintaining mechanisms is often considered as an obstacle significantly restricting application of reductionistic approaches. On the other hand, it is a blessing too, since with the right approaches such close interrelationships allow to naturally and unambiguously link workings of different organism's constituents, which facilitates understanding of an organism as a single system. Organism, indeed, *is* a single system, and so in one way or another one eventually has to come to systemic view of living creatures. The problem is to define the optimum path to this strategic goal, in which the general growth law may help.

Although we consider application of the general growth law to growth and finding metabolic characteristics of organs, and liver in particular, the introduced methods can be used for similar studies at other scale levels, such as cells or whole organisms. Since characteristics found this way at different dimensional scales are derived from the general growth law (through its mathematical representation the growth equation), this common foundation allows to naturally interrelate growth and metabolic characteristics at different scale levels within a single integral model.

The VLN project did not yet provide a common base for integrating different methods into a single framework. Similarly to the publications discussed above, presently it incorporates diverse approaches, such as kinetic, when one analyzes certain composition of biochemical reactions; creating models of biochemical signaling networks; stochastic modeling, when growth, replication and regeneration processes are considered rather on the basis of biomechanical properties of cells and tissues; vascular systems models, etc. We will consider in detail this project and publications relevant to our study in the second part of this paper that presents a general framework for modeling biological phenomena.

Ideally, the direction in which development of organ models should go is to create a *complete* model, in which all metabolites and intermediate products are considered, at all phases of cell, organ or organism cycle, from origin to fully grown size, including possible transition from the growth phase to grown and quiescent states. Such models would allow to better understand biochemical and morphological dynamics and provide solid foundation for reliable control of biological processes. Certain similarity can be found in biotechnological applications, where a clear trend to add more biochemical processes, from the major backbone reactions to inclusion of full genome information [12], is observed. With the increase of the models' size and complexity the issue of their stability and unambiguity may arise. Such issues should be addressed too.

## 3. Importance of Integral Constraints. Integral Nutrient Influxes

One of the major problems in present approaches to modeling metabolism of living organisms and their constituents are difficulties in finding the *actual* quantities of substances involved in biochemical reactions, and the need for robust and efficient methods to describe *dynamics* of biochemical reactions, that is their *continuous* transformation during the whole organisms' life cycles or certain phases [12]. Certainly, when we speak of continuity of transformation, we do not discard possibility of abrupt changes, but rather speak of possible significant differences in rates of changes in composition of biochemical reactions at different stages. Addressing these issues, and especially when the modeled phenomenon includes interrelated mechanisms acting at different dimensional scales, requires imposing *integral* constraints, such as the rate of nutrient consumption by an organism and its constituents. Creation of such integral models will require knowledge of distribution of nutrients between different activities (biomass synthesis, support of organism's maintenance needs, synthesis of proteins, RNAs, etc.) Nutrient consumption is the essence of organic life. This inherently "nutritional" (and consequently metabolic) nature of living organisms is that backbone which comprehensive biological models and studies should incorporate.

Of course, many other factors influence nutrient distribution. In this regard, mechanical forces certainly deserve attention. Work [13] provides a comprehensive and thoughtful review of the subject. Example of relatively recently introduced stochastic modeling, which also accounts for mechanical properties of cell aggregations, can be found in [14]. However, in organisms, mechanical properties of tissues and cell aggregations are not independent of the nutrients distribution and organism's metabolism, and vice versa. The fact that growth processes produce changes in mechanical forces confirms this. Thus, adequate models of biological phenomena based on mechanistic approaches generally have to directly or indirectly integrate metabolic characteristics.

Results of recent studies and particular applications of the general growth law presented in [1,12,15–19], with major work [1], allow finding metabolic parameters, which effectively define integral constraints required for complete models of biochemical mechanisms. Another important aspect is that these integral metabolic parameters, defined through the general growth law, are *dynamic* ones, and can be determined at every phase of growth and at each spatial level. This important feature of proposed methods lays a foundation for developing models that describe a truly continuous transformation of composition of biochemical reactions (and consequently the dynamically changing amount of metabolites and intermediate products) during organism life cycles.

## 4. Amount of Produced Biomass as a Leading Parameter

The general growth law and its mathematical representation, the growth equation, were introduced in [1]. For metabolic studies, it is important that this law *uniquely* defines distribution of nutrients between two major activities inherent to *any* living organism – biomass production and support of existing biomass, also called maintenance needs. The parameter, which mathematically represents this fundamental mechanism and quantitatively defines this division of nutrients, is the growth ratio. It depends on the geometrical form of a considered organism or its constituent, and indirectly on the properties of biochemical machinery. Given evolutionary pressures which required fast replication as a condition for survival in fierce competition for resources, such an optimization of resource distribution had to evolutionarily emerge. Indeed, too much nutrients for maintenance is not optimal, since in this case rate of biomass synthesis will be slow. On the other hand, if too much nutrients are used for biomass production, then maintenance needs will be impaired, which reduces survival chances too and suppresses biomass production. This arrangement could be understood better if we recall that cells and other organisms' components do not have separate sets of biochemical reactions, that is one specifically for maintenance and the other exclusively for biomass synthesis. In fact, both activities are supported *by the same and the only available* biochemical machinery, in which all reactions interrelate [12,20].

Since the growth equation uniquely defines how much nutrients go to biomass production, once we know the total influx of nutrients, we can determine *absolute* amount of nutrients required for supporting maintenance needs. Besides, for particular organisms, once we know their certain properties, we can further determine distribution of nutrients between downstream activities, such as nutrient influx for RNA, DNA, protein syntheses, etc. [1,12,15]. Note that due to conservation law of matter, we can say that the mass of nutrients used for biomass production will be equal to the mass of produced biomass, and the same is true for other synthesis processes. So, from the said above, the following arrangement emerges. Using the growth equation, we find amount of nutrients that goes to biomass production. When we know the total amount of nutrients (which can be found based on different considerations, including methods presented in [1,12,15]) and amount of nutrients that goes to biomass production, we can find amount of nutrients required for maintenance. Then, the composition of biochemical reactions should be such that it matches the amount of produced biomass *and* the amount of nutrients required for maintenance.

In order to prove the validity and efficiency of this approach, let us take a look at biotechnological applications which deal with biomass production using specific methods, in particular methods of metabolic flux analysis [21,22]. Biotechnological applications based on these methods show that amount of produced biomass is one of the major parameters that define composition of biochemical reactions. Indeed, methods of metabolic flux analysis (the main analytical approaches in biotechnological applications) produce the most adequate results when solution of system of stoichiometric equations is optimized for a *maximum amount* of produced biomass [21,22]. This is an expected result given the fact that evolutionary pressures generally force natural selection in the direction of faster reproduction. Accordingly, this means faster growth and consequently faster synthesis of biomass. Of course, this evolutionary optimization was a result of compromise between this selection requirement and other constraints imposed by the environment. However, for given characteristics of a developed organism (form, size, inherent biochemical machinery, etc.) the growth process was evolutionarily optimized for a maximum amount of produced biomass at *every phase* of growth and *at every* dimensional level. So, knowledge and experience gained in biotechnological applications confirm our assertion about the leading role of biomass production in defining composition of biochemical reactions. This leads us to the following generalization, which was founded and presented in detail in [1,12,15]: amount of produced biomass is a *leading* parameter that defines composition of biochemical reactions in all organisms. This conclusion is valid both for the evolutionary development of all living species, as well as for their individual growth, existence and replication on everyday basis. Figure 2 illustrates this arrangement graphically. (Generally, in Nature, the utility function of organisms varies, which affects optimization criteria. However, since organisms are composed of cells, and cells overwhelmingly were evolutionarily developing in the direction of optimizing biomass production (for many reasons), this means that growth of cells' aggregations generally should be also optimized for maximum production of biomass, both at cellular and higher scale levels.).

In the diagram, initial state (geometry and biochemical properties) of an organism or its constituent defines the value of the growth ratio, which accordingly defines how nutrient influx is divided between maintenance needs and biomass production. The *uniquely* defined nutrient influx for biomass production accordingly *uniquely* defines amount of produced biomass. Since the geometry of a growing entity changes, the value of the growth ratio (and accordingly the distribution of nutrients between biomass synthesis and maintenance) changes too. Composition of biochemical reactions is continuously adjusted to the changing amount of nutrient influx diverted to biomass production, which is equated to the amount of synthesized biomass according to the law of conservation of matter. Thus, constraints imposed by the general growth law (through the growth ratio) define amount of synthesized biomass, while biomass increase continuously, in a feedback manner, changes the size and geometry of a growing entity. Because of the size increase, more and more nutrients will be needed to support maintenance needs, so that eventually amount of nutrients will be sufficient to only support organism's maintenance needs and no nutrients will be available for growth. The lack of nutrients for biomass production will stop growth. This is how organisms and their constituents progress through their life cycles.

A note should be made about relationships of the "Composition of biochemical reactions" and "Maintenance" blocks in Figure 2

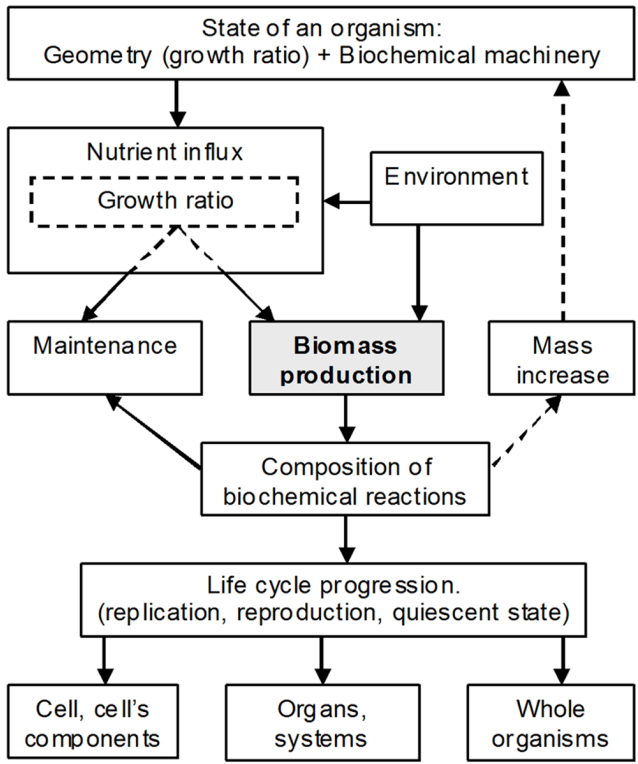

**Figure 2. The role of biomass production.** How the amount of synthesized biomass orchestrates and directs cooperative working of different growth and replication processes and factors during evolutionary development and individual life cycles.
doi:10.1371/journal.pone.0099836.g002

connected by a one-directional arrow toward "Maintenance" box. How well is this direction justified? Could maintenance needs influence the composition of biochemical reactions? We think that the answer to the last question is "no". Biomass production is of primary importance for survival of organisms. Maintenance *serves* this purpose, so it is a subjugated activity. Let us assume the opposite, that is maintenance is a primary activity. However, maintenance for the sake of maintenance does not have evolutionary sense, since it does not lead to a successful reproduction. Consequently, composition of biochemical reactions for maintenance part should also depend on the amount of produced biomass.

Note that the described mechanism, the general growth law, is also one of the major factors responsible for *balanced* growth of organs and systems in the same organism. Indeed, functioning of different organs and systems is adjusted to each other. Strong muscles are impossible without a strong heart. Such perfect balancing of different organs could be based only on some general mechanism which universally affects *all* organs and systems in *any* living organism. If it is a genetic mechanism, then it has to be the same across *all* living species, otherwise the phenomenon would not be *universal*. But it should not be the case given enormous variety of genetic configurations and arrangements. However, universal distribution of nutrients between growth and maintenance, indeed, could be such a solid foundation, since it is *universally* present in *all* living species. Such a mechanism then perfectly fits the task of balancing growth of different organs and systems, a task of enormous importance. In fact, if not for this universal mechanism that balances growth of different constituents of an organism, life would likely be impossible, except perhaps in its simplest forms.

Below, we show how to find metabolic characteristics of growing and existing organisms using the general growth law. For illustration purposes, we rely on the same experimental data from works [2] for dog livers and [3,4] for human livers. This application of the general growth law is unique in that regard that it allows interrelating general characteristics of organisms (mass, volume and geometrical parameters), directly to metabolic properties such as nutrient consumption rate and distribution of nutritional resources between the maintenance needs and biomass synthesis. Simplicity and elegancy of application of the general growth law to this important task makes the approach very effective.

## Methods

### 1. Finding Metabolic Characteristics from the Growth Equation

The growth equation depends on several parameters [1]. The aforementioned *growth ratio*, which uniquely defines the fraction of nutrients that goes to biomass production, depends on the geometric shape of the organ. Let us assume that nutrient availability and the biochemical specifics of an organ receiving nutrients through the surface, allow the organ to grow to a maximum volume of $V_{MAX}$ with a maximum surface of $S_{MAX} = S(V_{MAX})$. We define the *dimensionless* relative surface $R_S$ and the relative volume $R_V$ as $R_S = \dfrac{S(V)}{S(V_{MAX})}$, $R_V = \dfrac{V}{V_{MAX}}$. Then, the growth ratio $G_R$, which is also dimensionless, is defined as $G_R = \dfrac{R_S}{R_V} - 1$.

Although the growth ratio is described in terms of geometric characteristics, it is closely related to the biochemistry of the organ, since it defines how much nutritional resources are used for biomass production, while the rest supports the organ's maintenance needs. A particular form of the growth equation depends on the growth scenario. When nutrients are supplied through the surface, the growth equation can be written as:

$$p_c(X)dV(X,t) = \left( \int\limits_{S(X)} k(X,t) \times dS(X) \right) \times \left( \frac{R_S}{R_V} - 1 \right) dt.$$

Here, $X$ is the spatial coordinate, $p_c$ is the density of the tissue measured in $kg/m^3$, $t$ is time, $k$ is the specific influx, which is the nutrient influx per unit surface per unit time measured in $kg/(m^2 \times \sec)$, $p_c(X)dV(X,t)$ is the change in mass, and $dS(X)$ is the elementary surface area. When the specific influx does not depend on the location of an elementary surface area, it transforms to the following.

$$p_c(X)dV(X,t) = k(t) \times S \times \left( \frac{R_S}{R_V} - 1 \right) dt \qquad (1)$$

where $S$ is the total surface through which nutrients are supplied. This equation has the following interpretation: The left-hand side represents the mass increment. The right-hand side represents the total influx through the surface, that is the term $\int\limits_{S(X)} k(X,t) \times dS(X)$, multiplied by the growth ratio $(R_S/R_V - 1)$, so that this product defines the amount of nutrients that is available for biomass production. Now, using equation 1, we can find nutrient influxes per unit of volume required for maintenance $(K_M)$ and biomass synthesis $(K_B)$ as follows.

$$K_M(t) = k(t)S(t)(1 - G(t))/V(t) \qquad (2)$$

$$K_B(t) = k(t)S(t)G(t)/V(t) \qquad (3)$$

Appropriate cumulative amounts of nutrients for maintenance $M$ and for biomass synthesis $B$ for an arbitrary period $[t_1, t]$ can be found as

$$M(t) = \int_{t_1}^{t} K_M(\tau)d\tau \qquad (4)$$

$$B(t) = \int_{t_1}^{t} K_B(\tau)d\tau \qquad (5)$$

Probably because of the name, sometimes people assume that the growth equation can be used only for describing growth. In fact, it can be used for different purposes, The growth equation presents mathematical relationship of several parameters (which, in turn, can depend on other parameters). Which parameters are to be found depends on the problem context. For instance, one can use the growth equation for parameter inference, including statistical applications. Presence of the size of a grown organism in the growth equation as a parameter reflects on the fact that the final grown size generally cannot be known before the growth completes, because growth conditions can change. Suppose some organism started to grow in a nutritionally poor environment. If it would continue to grow this way, it destined to have a small grown size. However, if the nutritional environment became rich in nutrients, then the organism's grown mass can become bigger (and its shape can change as well), which was proved in experiments on cell growth presented in [23]. This kind of variability of the final size, indeed, is how living species grow in Nature, and the general form of the growth equation describes this phenomenon through the variable grown mass (this topic was discussed in detail in [1]). In many instances, the final size can be known from the beginning of growth, like in case of doubling of many types of cells when nutrient supply is stable - examples are in [1]. Often, the grown size can be evaluated based on preliminary information, like growth of an organ, when organ's mass is usually a well defined fraction of mass of the total organism.

The general growth law acts at different scale levels, from organelles and cells to whole organisms, like matryoshka (Russian nested dolls). Its working at lower levels, like cells, cooperates with its working at higher scale levels, for instance, organs or vascular systems. The general growth law optimally divides nutrient resources between maintenance needs and biomass synthesis at *each* scale level, thus providing robust constraints at each such level during the whole growth cycle. This uniquely defined at every phase of growth distribution of nutrients between growth and maintenance is that mechanism which secures stable and *balanced* growth of *all* of an organism's constituents at *all* scale levels.

General growth law and biochemical machinery are inherently tied together, supporting each other at all spatial levels, since amounts of nutrients diverted to biomass production and maintenance define composition of biochemical reactions. However, this composition is not rigidly defined but has a certain range of variability; it can change, but the distribution of nutrients between biomass synthesis and maintenance can remain the same. Certain biochemical factors, like growth factors, directly or indirectly, through local or more global feedback loops can interfere with the current biochemical arrangement at different levels, and change it within certain limits, thus changing the path of growth scenario. Such interference can occur at any scale level, from intracellular to the whole organism. For instance, angiogenesis stimulated by oxygen demands of tissues can be triggered via local feedback signaling loops [24], while mechanical forces may affect biochemical growth mechanisms (mechanotransduction) at larger scales, up to the whole organism. It was noted in [13]: "mechanical stress has also been shown to be a fundamental factor in the development of organs and embryos, from eyes ..., to the brain ... and the heart". However, despite such possible variations of growth scenarios, distribution of nutrient resources between growth and maintenance at all spatial levels is defined by the general growth law, and the biochemical machinery (including its signaling part) must obey restrictions imposed by the general growth law in the form of uniquely defined distribution of nutrients.

The described mechanism of cooperative and mutually dependent working of the general growth law and organisms' biochemical machinery is in good agreement with numerous studies confirming high degree of interdependency between different biochemical mechanisms and complexes. Researchers often speculate, what is the base of this interdependency? The answer is this. The general growth law defines amount of nutrients diverted to biomass synthesis and maintenance, while the overall biochemical machinery of organisms is a *single* mechanism, in which all chemical components interrelate through a *single* composition of transforming biochemical reactions tied to amount of produced biomass as a *leading* parameter. Because of the singularity of the whole biochemical machine, it means that all other biochemical reactions serving other functions, such as specific maintenance mechanisms, signaling, transporting, etc. are also tied to amount of produced biomass. These are two fundamental properties, or mechanisms, which underlie such marvelous working of highly reactive and sensitive, while stable and balanced at all levels biochemical machinery of living organisms. Since the growth law acts at different scale levels, this common base accordingly, quite logically, assumes existence of biochemical mechanisms acting at different scale levels, from local at organelles' and cells' levels to global. Indeed, existence of both global and local regulative mechanisms is presently supported by many studies. In particular, work [24] presents evidence that *local* mechanisms account for angiogenic network formation, besides more global factors.

## 2. Growth Equations for Absolute Growth

Unlike in work [15], which considered relative volumes, in order to find nutrient consumption rate, we should use the growth equation in *absolute* values,. For that purpose, the growth equation should be rewritten as follows.

$$pdV(r, d) = K \times V(r, d) \times \left( \frac{R_S}{R_V} - 1 \right) dt \qquad (6)$$

According to [25], transplanted livers do not resume their normal growth from the very beginning, but there is a transitional period when less hepatocytes are involved in proliferation than in normal growth. In such case of partial liver growth, the growth equation is as follows.

$$pdV(r,d) = K \times V_A \times \left(\frac{R_S}{R_V} - 1\right) dt \qquad (7)$$

Here, $K$ is nutrient influx which we define as amount of nutrients per one cubic centimeter of liver per day $(g/(cm^3 \times day))$; $V(r,d)$ and $V_A$ are current liver volume and liver volume involved in replication, measured in $cm^3$; density $p$ is measured in $(g/cm^3)$; time $t$ is measured in days. We also assume that density $p = 1.0 \, (g/cm^3)$, which is a well justified assumption for most soft animal tissues. For this density, we may speak interchangeably in terms of volume and weight. That is, numerically, results in cubic centimeters or in grams of nutrients per unit of liver volume per day are the same. However, since experimental data are presented in units of volume, we will use volume as a base measure.

Partial growth was considered in detail in previous studies on liver growth. Briefly, the approach is as follows. During partial growth, we distinguish the growing part of the liver (we call it the "active" part below) from the part of the liver that does not participate in regeneration (the "passive" part). We take into account that the passive part still requires nutrients for maintenance, but do not contribute to biomass production. Computationally, this means that at each integration time step we transfer an elementary volume from the passive part to the active part.

We model the reduction of the passive part $V_P$ during growth as:

$$V_P = V_b(1-A) \times \left(\frac{V_J V_b - V_C}{V_J V_b - V_b}\right)^p \qquad (8)$$

where $A$ is the fraction of the initial active part, $V_b$ is the initial liver volume; $V_J$ is the relative volume at the joining point, when the normal growth resumes, $V_C$ is the total volume of the growing liver, $p$ is a power that allows varying the functional dependence $V_P(V_C)$, choosing different concave and convex shapes. Equation 10 reflects the monotonic increase of the active liver volume. The exponent $p$ accounts for deviations from purely linear increase (when $p=1$). Both parameters $p$ and $A$ are found by fitting to experimental data, including the one before the "joining point". The volumes at the joining points for dog 1 and dog 2, according to previous computations are $V_{J1} = 1.1576$ and $V_{J2} = 1.6918$. Note that equation 10 is constructed in such a way that the passive volume becomes zero when $V_C = V_J V_b$. The active growing part is the complement of the passive part, hence $V_A = V_C - V_P$.

The values of $A$ and $p$ are unique and robust. The reason is that parameter $A$ defines the shape of the *whole* growth curve, while parameter $p$ affects only the shape of the growth curve *before* the joining point.

The value of $K$ is the only free parameter in equation 6. We start from an arbitrary value, say $K=1 \, g/(cm^3 \times day)$, and use the following procedure to find the value of $K$ that corresponds to the data. Let us denote a set of experimental points as $\{E_i(V_i, t_i)\}$, and a set of corresponding points on the computed growth curve as $\{C_i(V_i, t_{Ci})\}$, where, in general, $t_{Ci} \neq t_i$. Then, we can find the average scaling coefficient $T_{AV}$, by which all abscissa coordinates of the computed growth curve have to be multiplied in order to match experimental data, as follows.

$$T_{AV} = \left(\sum_i w_i(t_i/t_{Ci})\right) / \sum_i w_i \qquad (9)$$

where $w_i$ is a weighting coefficient. In a simple scenario, these weighting coefficients can be assumed equal to one. In our case, we used the following approach: $w_i = (V_i/V_b)^p$. Since the larger divergence corresponded to larger volume, we used $p=3$ in order to give more weight to points with greater volume. Then, depending on whether $T_{AV}$ is greater than or less than 1, we decrease or increase the value of $K$. We repeat these two steps iteratively until $T_{AV} \approx 1.0$, that is when the computed growth curve coincides with experimental data. (Note that this procedure does not change the shape of the computed growth curve, but only scales it along the time axis.) Thus found value of $K$ corresponds to actual nutrient influx. Such a scaling procedure is very similar to classic non-linear regression and shares many of its properties. In particular, value $T_{AV}$ is uniquely defined, which, accordingly, leads to uniquely defined value of $K$ corresponding to actual nutrient influx. Then, using the value of the growth ratio, we can find nutrient influxes accordingly for maintenance and for biomass synthesis.

$$K_M(t) = K(1 - G(t)); \ K_B(t) = KG(t)$$

Appropriate cumulative amounts of nutrients for maintenance $M$ and biomass synthesis $B$, for an arbitrary period $[t_1, t]$, can be found as follows.

$$M(t) = \int_{t_1}^{t} K_M(\tau)d\tau; \ B(t) = \int_{t_1}^{t} K_B(\tau)d\tau$$

## 3. Finding Metabolic Characteristics for Livers Transplanted to Large Dogs

Here, we use the shape of the computed growth curve from the previous studies on liver growth (the article by Yu. K. Shestopaloff and Ivo F. Sbalzarini "A method for modeling growth of organs and transplants based on the general growth law: Application to the liver in dogs and humans", presently accepted for publication in PLOS ONE). The liver is represented by a partial torus with hemisphere caps at ends, all sliced in plane of symmetry. Since in our case the small dog livers at the beginning exhibit partial growth, we have to use the growth equation 7 for partial growth. Dimensions of dogs' livers models corresponding to volume of transplanted livers (in $cm^3$) and other parameters required for the growth equation are shown in Table 1.

Using data from Table 1, we applied the procedure for finding nutrient influx, described above, to the liver growth for first dog. The determined value of nutrient influx was $K=1.01 \, g/(cm^3 \times day)$. (Note that this an *absolute* value of nutrient influx, not a normalized one. It just happened that it is close to one for a particular dog.) In other words, the liver of this dog consumes about the same amount of food per day as the liver volume. For a grown liver, this is the amount of nutrients required for maintenance. Such a high value of consumed by liver nutrients means very active liver metabolism. This explains why liver in

**Table 1.** Dimensions of geometrical models of dogs' livers (absolute liver size).

| Parameter | Dog 1 | Dog 2 |
|---|---|---|
| Initial torus radius $r_b$ (cm) | 3.38783 | 3.29636 |
| Ending torus radius $r_e$ (cm) | 4.37317 | 4.778 |
| Initial distance from the center to torus axis $d_b$ (cm) | 4.23479 | 4.12045 |
| Ending distance from torus center to torus axis $d_e$ (cm) | 5.4665 | 5.9725 |
| Fraction of the torus used for modeling | 2/3 | 2/3 |
| Minimum initial volume (cubic centimeters) | 374.28 | 344.778 |
| Ending volume (cubic centimeters) | 805.05 | 1049.963 |
| Relative volume (relative to minimum volume) | 2.1509 | 3.0453 |
| Part of liver that grows at the beginning (relative to the whole initial liver volume), computed parameter $A$ in equation 8. | 0.67 | 0.51 |
| Computed power $p$ in equation (8), defining convexity of curve | 0.85 | 0.83 |
| Nutrient influx [g/(cm$^3$*day)] | 1.01 | 0.624 |

doi:10.1371/journal.pone.0099836.t001

many mammals grows so fast and probably why they can sustain many negative impacts, such as bad food, and also why they have such enormous digestive abilities (wolves digest bones). Higher body temperature of dogs also contributes to accelerated metabolism.

During growth, nutrient influx is distributed between maintenance needs and biomass production according to the value of the growth ratio, as shown in Figure 3. At the beginning, when the liver is small, more resources go to biomass production. When the liver grows, less and less resources are available for biomass synthesis, because more nutrients are required for maintenance of existing liver. Finally, nutrient influx becomes sufficient to only support liver's maintenance needs.

Obtained result can be validated as follows. Table 2 shows experimental data for dogs 1 and 2. We can see from Table 2 that during the third day liver volume increased by approximately $\Delta V_3 \approx (532 - 433.26) = 98.74 \ cm^3$. If our computations are valid, then we should be able to find biomass increase as where $G_{3AV}$ is an average value of the growth ratio during the third day; $V_{3AV}$ is an average volume during the third day; $D$ is the number of days. Substituting into equation 10 appropriate values found from calculations, we obtain $D \ cm^3$, which is very close to the actual change of volume of 98.74 $cm^3$. So, the found value of $K$ sustained this validation.

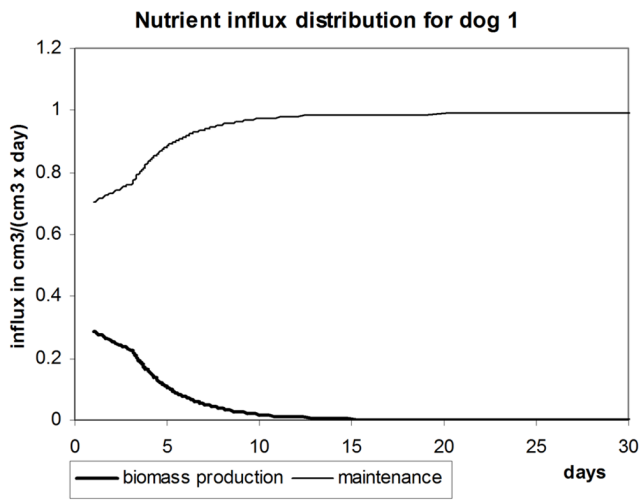

**Figure 3. Distribution of nutrient influx.** How much nutrients go to liver maintenance needs and to biomass synthesis depending on time for dog 1. Nutrient influx $K = 1.01 \ g/(cm^3 \times day)$.
doi:10.1371/journal.pone.0099836.g003

Note that such a validation is not a "self-fulfilling prophecy", since all parameters, except for the tested value $K$, are objectively defined by geometrical characteristics of liver (i.e. by the growth ratio) and experimental data (volume, number of days). It turned out that the validation criterion defined by equation 10 is a very sensitive *integral* measure of both the adequacy of a particular model and accuracy of experimental data. If this criterion is not satisfied, then either the model is not adequate, or data are inaccurate, or both. On the other hand, equation 10 can be used as an independent tool for finding metabolic rates for biomass production and maintenance, which we will demonstrate later in studies of human livers.

Dynamics of total nutrient consumption in cubic centimeters per cubic centimeter of liver depending on growth time is presented in Figure 4. In the limit, the line for cumulative nutrients required for maintenance quickly approaches linear asymptote. The curve that corresponds to nutrients that go to biomass production approaches horizontal asymptote. In real conditions, hepatocytes are replaced on ongoing basis, although very slowly, so that biomass synthesis and consequently nutrient consumption for biomass production is a continuous process.

Using the same approach, we found that the nutrient influx for the second dog was less ($cm^3 \ cm^3$). Given very likely differences in age, sex [26], which affect metabolism, this variation is justified. Besides, such a noticeable difference might explain certain effects. For instance, it is very likely that the first dog, with higher metabolism of liver, was female, in order to support birth function, as we will find later for humans.

The graph presenting distribution of nutrient influx between maintenance and biomass synthesis for dog 2 is presented in Figure 5. Compared to the first dog, we can observe higher proportion of nutrients diverted to biomass synthesis at the beginning. This happens because the original liver is smaller relative to the grown size than in case of the first dog. Consequently, the value of the growth ratio is higher as well, which accordingly increases relative amount of nutrients that go to biomass production.

Verification of the obtained nutrient influx for the second dog by experimental data can be done using equation 10. In this case,

**Table 2.** Liver growth in dogs.

| Days, dog 1 | Volume, cm³ dog 1 | Days, dog 2 | Volume, cm³ dog 2 |
|---|---|---|---|
| 0 | 438.3846 | 0 | 344.7778 |
| 1 | 374.2821 | 2 | 407 |
| 3 | 433.2564 | 3 | 456.2593 |
| 4 | 531.9744 | 4 | 504.2222 |
| 6 | 640.9487 | 6 | 583.2963 |
| 8 | 715.3077 | 8 | 725.8889 |
| 12 | 784.5385 | 12 | 894.4074 |
| 30 | 805.0513 | 30 | 1049.963 |

doi:10.1371/journal.pone.0099836.t002

we verified volume increase between 6-th and 8-th days. The actual volume change was about $\Delta V_{6-8} \approx 142.6 \ cm^3$, while computations yield very close value of

$$\Delta V_{6-8} \approx 0.169 \times 670 \times 0.624 \times 2 = 141.3 \ cm^3$$

So, the obtained value of $K = 0.624 \ g/(cm^3 \times day)$ sustained the validation test too.

Although we considered experimental data only for two dogs, nonetheless, the obtained results suggest that dogs have high level of metabolism both for maintenance and liver growth and regeneration, such, that the amount of food consumed by a dog liver per day is of the order of the liver weight. We also found that liver metabolism in different dogs may have noticeable variations. One of the likely reasons is difference in sex.

## Metabolic Characteristics of Human Livers

In this section, we study liver metabolism of male and female patients, donors of the right lobe grafts, using results from [3]. We use the form of the growth equation for absolute values defined by equation 6. Since the study [3] provided only average values for the representative groups of male and female patients, we created hypothetical average male and female livers using average value of the initial liver volume of 820 $cm^3$ for male patients. For female patients, whose liver grew less, we considered two scenarios: one is when the ending volume is adjusted to the ending volume for males (the group "Female E"), and the other when the beginning volume is adjusted to the beginning volume for males (the group "Female B"). Intermediate values were obtained by multiplying absolute initial volumes by a known relative increase [3]. Table 3 presents these datasets.

The human liver model was developed in the previous article about modeling growth of livers. Model parameters corresponding to volumes in Table 3, as well as other parameters required for computations, are shown in Table 4. Note that in case of human livers we need to know the final liver size and shape, since we consider grafts, but not entire livers, like for dogs.

Using these input parameters and applying proposed method for finding nutrient influx, we obtained $K = 0.033 \ g/(cm^3 \times day)$ for male patients, and noticeably higher value of $K = 0.044 \ g/(cm^3 \times day)$ for female patients. However, none of these values

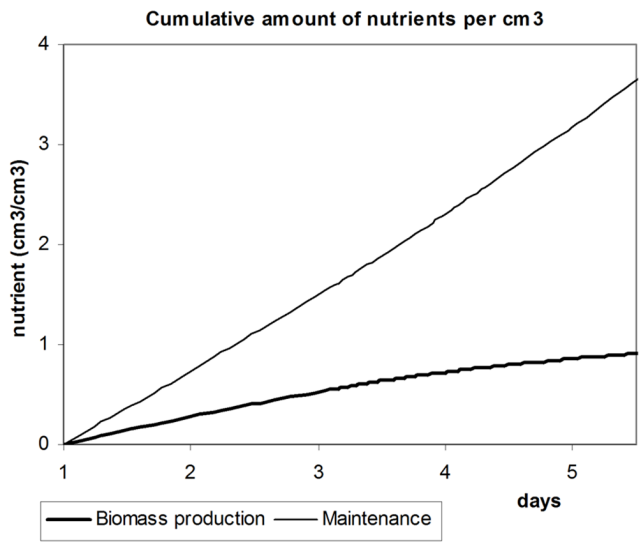

**Figure 4. Cumulative amounts of nutrients going to maintenance needs and biomass production.** Cumulative amount of nutrients consumed by the liver of dog 1 for maintenance and biomass production during liver regeneration, depending on time.
doi:10.1371/journal.pone.0099836.g004

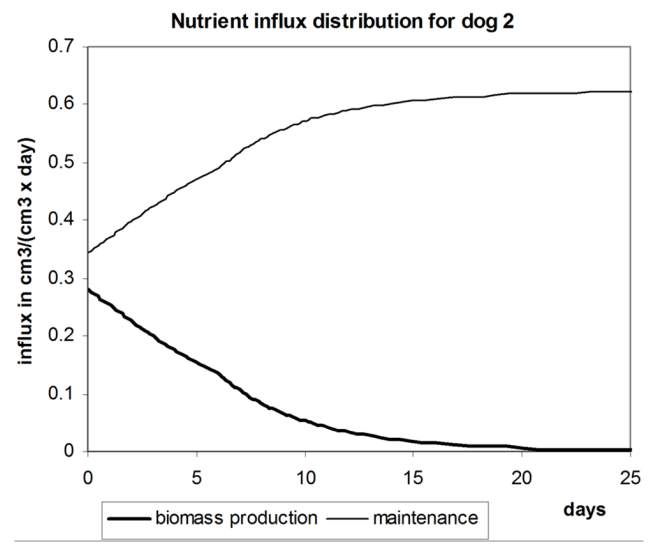

**Figure 5. Cumulative amounts of nutrients going to maintenance needs and biomass production.** Distribution of nutrient influx $K$ between liver maintenance and biomass synthesis depending on time for dog 2. $K = 0.624 \ K = 0.624$.
doi:10.1371/journal.pone.0099836.g005

**Table 3.** Growth of left lobe liver grafts in different hypothetical liver donors (cm$^3$).

| Days | Male | Female E | Female B |
|------|------|----------|----------|
| 0 | 820 | 925.4703 | 820 |
| 7 | 1049.656 | 1081.606 | 958.3415 |
| 30 | 1192.663 | 1250.973 | 1108.407 |
| 90 | 1278.693 | 1321.971 | 1171.314 |
| 180 | 1370.189 | 1417.693 | 1256.127 |
| 360 | 1444.158 | 1444.158 | 1279.575 |

doi:10.1371/journal.pone.0099836.t003

sustained the validation test by equation 10. There are at least two reasons for that. One is that most likely livers grew partially at the beginning, but there is no way neither to prove nor reject this assumption on the basis of available data. The other very likely reason is that data in [3] were compiled as statistical average for 27 male patients and 16 female patients with large divergence in the values of beginning volumes ($49.4 \pm 5.7\%$), ending volumes ($83.4 \pm 9\%$), remnant volumes ($35.1 \div 62.5\%$), total liver volumes ($1153 \div 2838 cm^3$), and patients' age ($36 \pm 9.6$) years. Besides, the growth equation should be used for a *single* growing entity (a cell, an organ or an organism), while in our case the analyzed data represent statistically non-uniform groups. So, the application of the growth equation to the data sets that represent groups, and at the same time have large diversion from average, is unlikely to produce accurate information about nutrient influx. For that purpose, we would need more accurate data for a *single* organism (or an organ, or cell). In this regard, accuracy of experimental results for dogs, when we were able to accurately estimate the nutrients influx and validate results, can be considered as a benchmark.

We also studied the influence of variations of a geometrical form, "thickening" the liver by up to two times (which is an extreme value), but this did not influence results much, leading to an increase of nutrient influx of about 10%.

On the other hand, we can assume that the liver remnants grew partially at the beginning, which is a reasonable assumption given results obtained earlier for the dog livers. There were strong indications presented in [25] that human liver grafts switch to normal regeneration after 6–7 days. In this case, we can find the value of nutrient influx using equation 10. (Such calculations are possible since we know the volume increase, the growth ratio, the number of days and the average volume, so that the only unknown parameter in equation 10 is the nutrient influx $K$.) This approach yields the following: $K = 0.088 g/(cm^3 \times day)$ for males, and a higher value of $K = 0.098 g/(cm^3 \times day)$ for females. Sexual differences in metabolic rates should not be of surprise, given many evidences of substantial differences in functioning of male and female cells [26].

The authors note in [3] that "Female donors had significantly slower liver regrowth when compared to males at 12 months ($79.8 \pm 9.3\%$ versus $85.6 \pm 8.2\%$)". This result is well explained by the discovered higher metabolic capacity of female livers, which is most likely due to the need to support pregnancy. So, neither female liver transplants nor liver remnants in females need to grow as big as in males, since their higher metabolic capacity allows supporting metabolic requirements by having smaller size. This, accordingly, results in smaller final size of transplanted and regrown livers in females. Such a sexual distinction is an important factor to be taken into account in clinical practice. For instance, it means that a female donor can be safely left with a smaller part of liver than a male donor, while for the male donors the size of liver remnants is more critical for successful recovery, all other factors being equal.

Although we did not succeed in validation of the first set of results for human livers, the obtained values of nutrient influx at least give an idea about the magnitude of nutrient consumed by human livers to support their regeneration and maintenance. Apparently, metabolism in human livers is several times slower than in dogs, which seems as a reasonable result given significant dietary and environmental differences. The results also suggest that the liver metabolism in females is noticeably higher than in males, which can be explained by the fact that during pregnancy female's liver in addition has to support child's organism. If we assume that such an additional liver load is proportional to the total relative weight of a pregnant woman (which is a reasonable assumption), then we should expect, at least, 7–10% higher metabolic rate in female livers, which is on par with our result of about 11% (0.098 for females versus 0.088 for males). This result, if confirmed, would have many interesting implications.

**Table 4.** Input parameters for a human liver model for absolute volume calculations.

| Parameter | Male | Female E | Female B |
|-----------|------|----------|----------|
| Width of a grown liver (relative units) | 6.048922 | 6.048922 | 5.80981 |
| Small base $B$ of a grown liver (in units of width) | 6.048922 | 6.048922 | 5.80981 |
| Large base of a grown liver $B_x$ (in units of width) | 21.171227 | 21.171227 | 20.33433 |
| Length of a grown liver (in units of width) | 17.541874 | 17.541874 | 16.8485 |
| Relative initial liver volume | 0.5678 | 0.64084 | 0.64084 |
| Nutrient influx $K$ [g/(cm$^3$*day)] | 0.033 | 0.044 | 0.044 |

doi:10.1371/journal.pone.0099836.t004

## Analysis of Existing Biological Models

Framework is the next level of generalization to which many disciplines evolve once a certain combination of particular hypotheses, methods, approaches, concepts, methodologies, tools and other cognitive instruments has been established within the discipline. The crown of scientific truth is a *theory*, which is the best available proven knowledge that was thoroughly verified and accepted by the scientific community. In this regard, a framework is rather a well defined toolset consisting of interrelated general concepts and associated with them methods and general constraints. Together, these factors define the boundaries of areas of application and strategies to be employed. Ideally, they should also define robust and complete set of validation criteria.

Frameworks can change by introducing more newer and proven advanced concepts, methods, constraints, etc., while the valid scientific theory cannot reject valid older concepts but must incorporate them as a particular case [27]. In such disciplines like classical physics the term "framework" is not used. However, such frameworks nonetheless exist and routinely used. Unlike in less developed, or more descriptive, or more complex disciplines composed of diverse loosely related constituents, such implicit frameworks are usually based on fundamental laws of Nature. The laws are presented by mathematical apparatuses, whose generality and completeness are sufficient to consistently and efficiently handle discipline's problems within such naturally developed frameworks. For creating a framework for biological modeling, we will exploit such implicit exemplary frameworks used for solving problems in optics [28], electronics [29], remote sensing [30], electromagnetic theory [31]. Some useful insights can be gained considering *explicit* development frameworks, which exist in more abstract disciplines, such as applied and financial mathematics [32]. The last work deserves attention because it defines a *complete* set of constraints and validation criteria required for development of models. The software system design is another discipline which provides valuable insights how to structure development frameworks and arrange their components [33].

Biology is a complex discipline consisting of many diverse parts, many of which employ specific methods and concepts that apparently are not transferable to adjoining areas. The issue is also complicated by a wide range of goals pursued by different biological studies. So, creating a general all purpose modeling framework in biology is a big challenge. In our study, we first introduce such a general modeling framework, and then analyze to what extent it can be used for modeling livers.

## 1. Virtual Liver Network (VLN) Project and Examples of Integration Approaches

*The goal of VLN project* is "to demonstrate an impact on the needs of patients and clinical practitioners alike". The general strategy for achieving this goal was formulated as follows: "…integrate the vast wealth of detailed information we have acquired from the advances in molecular and cellular biology since the completion of the human genome programme, not just in a mathematical model, but more importantly in a series of models that are linked <u>across scales</u> to represent and simulate organ function. This programme is prototyping ways to achieve true multi-scale modeling within a single organ and linking this to human physiology".

Note that the notion of "linked models" implicitly assumes connecting models through their outer interfaces. In fact, comprehensive integral models and systems usually have more sophisticated structures that could include parallel branches, trees, diverse types of hierarchies and associations, inclusions, when some models are internal components of others, etc. [33]. So, one of the important issues in creation of integral models and systems is to define their overall structure and associated workflow. The sooner this work starts, the smoother such projects go. It does not mean that the initial structure is fixed from the beginning. The development, or any other nontrivial undertaking for that sake, is an inherently *iterative* and *incremental* process, at all levels. This is the most natural and the most productive way to maximize progress. During the development the structure is inevitably subject to changes and adjustments as any other model's constituent is. The thing is that such iterative and incremental design process has to *converge*; ideally, to an *optimum* solution [27,33]. So, instead of linked models the VLN project's strategy should be rather directed towards developing *efficient optimum structure* of an integral model and its workflow, thus explicitly emphasizing importance and high priority of this issue.

**1.1. Integration of different models.** Although some methods for integration of different biochemical mechanisms within a single model were developed, by and large, the concepts and mechanisms that could reliably and universally unite different approaches to create a single coherent live model are yet insufficiently developed. The reason is low compatibility of used methods and approaches. In [34], the authors acknowledge, "Adequate analytical methods for a systemic consideration of the underlying processes are still missing. However, such multiscale approaches are necessary to understand the highly complex and intertwined structure of biological networks and the interplay with the surrounding organism". The work itself "present (s) an approach relying on dynamic flux balance analysis that allows the integration of metabolic networks at the cellular scale into standardized physiologically-based pharmacokinetic models at the whole-body level." The proposed method can be classified as a "vertical" integration, when from the lower cellular level the model allows to elevate to organ and organism's scales. (Another type of integration can be called "horizontal", when integrated methods and/or approaches are at the same spatial level.).

The approach proposed in [34] was tested by integration of "a genome-scale network reconstruction of a human hepatocyte into the liver tissue of a physiologically-based pharmacokinetic model of a human adult." Similarly to our considerations, the authors rightly note that "metabolic network needs to be considered within the context of the surrounding tissue and organism since the various levels of biological organization are mutually influencing each other." The model allows finding metabolite concentration profiles in the body and the surrounding liver tissue, and it was "applied to three case studies covering fundamental aspects of medicine and pharmacology: drug administration, biomarker identification and drug-induced toxication." [34].

**1.2. "Vertical" integration.** In order to do a "vertical" integration, the authors of [34] used case-specific objectives such as "maximization of ammonia production or maximization of uric acid production to quantify extracellular exchange rates with regard to a specific set of boundary conditions", which allowed obtaining *quantitative* metabolic characteristics. As the authors say, "In contrast, the underlying flux space is assessed qualitatively." However, it does not prevent from obtaining quantitative characteristics through several-step iterative procedure. In short, the main idea of this integration method is to combine stoichiometric metabolic networks at the cellular level and "ODE-based, physiologically-based pharmacokinetic … modeling". These two methods are integrated via metabolites exchange rates and metabolic flux, which are associated with nutrient influx and drug concentration. The drug interferes into the composition of biochemical reactions in the same way as we described at the beginning of "Methods" section considering the working of the general growth law.

The described "vertical" integration exploits mechanisms based on flux analysis. However, one set of parameters in the growth equation also relates to nutrient influx. Since the general growth law universally works at all scale levels, it means that fluxes at all these levels naturally relate to each other through the growth equation, which thus creates a common base for quantification of fluxes. This allows to potentially use the general growth law for solving one of the major problems of the discussed method, which the authors state as follows: "intracellular flux distributions of biological relevance can hardly be identified using these functional objectives since they rather evaluate the macroscopic behavior of the cell". The general growth law is that instrument which allows doing exactly that. This is possible because the general growth law allows finding nutrient distribution at *each* spatial level and then substrates' profiles (through flux balance analysis). In fact, the situation is even better than that, because application of the general growth law in many instances allows to significantly simplify the process of finding metabolites concentrations through the introduction of additional constraints [12]. Note that these constraints can be *dynamically* defined. So, by combining the general growth law and the discussed method, advanced integral models can be developed.

Another example of a "vertical" integral model can be found in [35]. It combines kinetic models representing the dynamics at a cellular scale, a two-dimensional liver model, and whole body model. Global reconstruction of human metabolism was done in [36]. The model comprehensively describes biochemical metabolic networks and metabolites. For our purposes, it is important to note that these models, capable of producing important practical results, are based on *biochemical* approach.

**1.3. "Horizontal" integration.** Review [37] presents examples of "horizontal" integration. Models integrate signaling, gene regulatory, and metabolic networks. These networks used to be considered as separate phenomena despite their close interdependency. In most instances, the approaches and formal mathematical apparatuses that are used to study these networks are also different. The study focuses mostly on models that are based on networks generally described by hypergraphs, which is a logical choice given the necessity to model chemical interdependencies between different substrates. The authors of [37] describe the situation as follows: "On the one hand, ordinary differential equations (ODEs) describing the underlying biochemistry are often used, as they are detailed and have high explanatory power. However, their applicability is limited due to the difficulty to obtain the necessary model parameters. They also have limited scalability, and thus they are, in general, not applicable to genome-scale models and simulations. On the other hand, less detailed approaches like Boolean networks and constraint-based models have been used in larger networks. Choosing the best modelling formalism is a trade-off between detail and complexity".

In our view, the choice is not as limited, since more details (which suggest more comprehensive knowledge or more efficient tools and methods) generally do not necessarily mean higher complexity, as the history of science and technology shows [38]. It is rather a matter of discovering new *qualitatively* more efficient approaches, tools, concepts, methods, etc. The best solution method is the one that is invented *for the problem*, because all other methods were created for *different* problems, and so they *fundamentally* cannot be as efficient for solving a "foreign" problem as the method that is exactly "tailored" to it.

*Signaling networks* are described by different types of mathematical formalisms. One class of approaches can be called "biochem-ical description", which is based on the chemical reactions underlying signal transduction. The other group is presented by "causal description" methods, where connections between network nodes are formally described without taking into consideration molecular details (like Boolean networks). Summary of features of both approaches can be found in [37].

Biochemical approach is usually more complicated, while the "casual" provides a greater flexibility. Both approaches have benefits. Which one to choose, depends on the efficiency of an approach relative to the task and *verifiability* of obtained results. With regard to verifiability in a broad sense (meaning criteria of validation of scientific hypotheses), the major validation criterion of hypotheses is *practice*, which in our case can be understood as an active interaction with objective reality and receiving feedbacks to our actions. These feedbacks are compared to our expectations and predictions (for instance, inferred from models). Depending on the result of comparison, a conclusion about the validity of models, or any hypothesis in general, is made. The closer some model reflects objective reality (in other words, natural features of considered phenomena, such as physical or chemical ones), the easier the model's validation is. The more intermediate abstraction layers a model introduces, the more difficult such validation will be, because of the necessity to transform the description of objective reality through abstraction layers in order to do validation. Such transformations inevitably distort description of natural features of the phenomenon.

*Gene regulatory networks* control gene expression. The authors of [37] evaluate the present situation with these models as follows: "Popular methods to reconstruct these networks include Bayesian inference, approaches based on mutual information, and modular approaches to reduce the problem's complexity. These have resulted in a few genome-scale models, mostly restricted to transcriptional regulatory networks…, as well as numerous other networks of small/medium scale for sub-systems of interest in biomedical research. Similarly to the signal transduction networks, formalisms for representing regulatory networks range from Boolean approaches for larger-scale networks to ODEs for small/medium-sized networks.".

*Metabolic networks* are different from gene regulatory and signaling networks, because they generate *mass* flows. So, logical or Boolean networks cannot be used for their description. Large-scale metabolic networks are represented as stoichiometric reversible biochemical networks. In essence, such networks are based on steady states. However, through introduction of certain enhancements they can reflect on dynamics through "pseudo steady-state conditions". In essence, these are constraint-based methods, "which include flux balance analysis, metabolic flux analysis, pathway analysis by elementary modes or extreme pathways" [37].

Note that successful *genome-scale metabolic models* were reconstructed and verified by methods developed for metabolic networks. Besides, the same methods were successfully used in metabolic engineering and biomedical issues related to cancer mechanisms.

So, the constraint-based methods provide very powerful approach to model metabolic networks. Note that these methods introduce little abstraction but rather deal directly with real *measurable* physical and chemical values, which significantly facilitates their verification and following improvement, as well as provides models' transparency and stability. On the opposite side, the authors of review [37] note that "Simulation of large metabolic models requires a huge computational effort, therefore model reduction is often used to reduce the size of the model and consequently the complexity of the mathematical problem". The aforementioned "huge computational effort" is due to algorithmic procedures required for finding solution. These procedures are based on certain constraints and optimization criteria. In fact, introducing appropriate additional constraints, such as amount of produced biomass, which can be found by independent methods, for instance, using the general growth law [12], will *significantly* reduce computational overhead. This can be done in a way similar to what we discussed for "vertical" integration, referring to original work [12], in which such an approach is introduced and demonstrated using example of *S. cerevisiae*.

So, although signaling, gene regulatory and metabolic networks can use similar mathematical methods based on ODEs, their integration on this basis is difficult for several reasons. One is different time scales of processes within each class of networks, "from seconds to a few minutes for signaling and metabolism to hours for gene regulation", [37]. The other reason is that the required input information is often difficult if not impossible to obtain. So, the authors think that "it is not expected that fully mechanistic ODE models integrating all layers will become available in the near future".

They see the situation as slightly more favorable with regard to combining logical models for signaling and gene regulatory networks, but different time scales and lack of knowledge about molecular interfaces between these networks, and the difficulty obtaining required data make such integration questionable too.

There were attempts to combine networks in *pairs*, like signaling and gene regulation, signaling and metabolism, and gene regulatory and metabolic networks. For instance, in one method, "regulatory Boolean network specifies the set of "inactive" enzymes for a given environment and a cellular state. This information is then used to constrain the fluxes of the respective reactions in the metabolic layer, simulated using FBA" (flux balance analysis) [37].

Additional studies resulted in two models that unite all three networks in a cell using different mathematical representations of each layer. The authors of [37] consider this development rather as "a proof of concept showing that all the processes in a cell can be modeled in an integrated fashion, assuming different time-scales of operation".

The authors formulated three computational aspects to be addressed in order to integrate different types of biological networks: "(i) a mathematical formalism has to be adopted to represent each layer; (ii) a simulation method capable of accounting for the different types of systems has to be developed; and (iii) the interactions among different layers have to be identified and modeled". Note that the authors, similar to "linked models" strategy of VLN project, which we discussed before, do not consider possibility of more elaborate structure of the integral model. Maybe more inclusive models with hierarchical relationships between components could provide better results.

Fulfilling the above requirements is difficult when one employs heterogeneous formal and phenomenological approaches, which is the case of considered networks. The situation could be improved if one finds a common base for these networks, especially if such a common base is founded on their *natural* unity, through which their integration (possibly with inclusion and more appropriate structure) could be facilitated. These networks coexist and cooperatively work within the same living organism, so that *there is* a common base for them. In this regard, the general growth law is a good candidate for this role, although not necessarily an immediate one. In any case, the influence of the general growth law on both the evolutionary development and life cycle of individual organisms and its constituents, in particular with regards to signaling, gene regulatory and metabolic networks, is too critical to ignore it [1].

## 2. Models of Vascular Systems

We already mentioned the work [24] which uses "a cell-based mathematical modeling approach (lattice-gas cellular automaton)". It belongs to the class of *stochastic simulation models* we will consider in detail later. Note that such models include many parameters, some of which are difficult to define. Overall, as we will see from the following analysis, despite certain useful features, such models are difficult to integrate with other approaches.

Work [39] presents overview of related works and introduces an algorithmic "concept for extending measured vascular tree data" through specific topological "ordering of the vascular trees, statistical testing, and averaging". The overall framework was used to discover additional details in a patient-specific hepatic vascular data. Authors suggest that their approach can be used in studies of other tissues.

Works [40,41] introduce models of hepatocytes growth and prove that hepatocytes align along the liver sinusoids (blood microvessels): "cells tend to arrange themselves in columns oriented towards the central vein", [40]. The result is in good agreement with the general growth law. According to growth equation (equation 1), the rate of growth is proportional to nutrient influx $k$. So, the closer hepatocytes are located to nutrients, the faster they grow. However, nutrients are supplied by the blood flow through sinusoids, and so the growing hepatocytes should tend to align along sinusoids, with prevailing direction of growth to higher concentration of nutrients, which is the central vein. Indeed, in [40] the authors confirm that "within the liver, lobule cells migrate preferentially into the direction of the central vein triggered by cell–cell interaction forces".

This "packaging" alignment has to be *optimal*, that is to simultaneously satisfy several optimization criteria, such as maximizing the number of hepatocytes for their optimal *orientation* and optimum *blood flow*. The last condition, in turn, presents maximum "interception" of the blood stream by individual hepatocytes, and a maximum "interception" of the overall blood stream by all hepatocytes on the way from the central vein to periportal veins. (The liver's main function is maintaining organism's homeostasis by processing the blood substances, and for that the liver has to process the *whole* blood stream.) Another factor, the *maximum throughput* of the blood flow through liver should be taken into account as well, for which the hydraulic resistance should be minimal.

Thus, from a simple consideration based on the general growth law, we first made a conclusion that hepatocytes should grow in a maximum proximity to sinusoids, and then formulated the problem of hepatocytes growth in damaged livers and defined a set of optimization criteria.

In particular, it is intuitively clear that hepatocytes formations satisfying the above requirements have to be compact (criterion of maximum possible number of hepatocytes), provide maximum contact area with the source of blood supply (sinusoids), and have relatively smooth surface (in order to minimize hydraulic resistance). One-cell-width sheets of densely packed hepatocytes, which are oriented in certain way, satisfy these requirements. Actually, this is how the hepatocytes are arranged in livers. These inferences agree with the results obtained in [40]: "coordinated cell orientation and cell polarity were identified to be the most critical parameters". Indeed, lack of coordinated cell orientation and polarity immediately leads to difficulties in satisfying introduced requirements. Such, compact "packaging" of hepatocytes actually becomes impossible, formation of smooth surfaces composed of "packaged" hepatocytes becomes very difficult too, so that the hydraulic resistance significantly increases, and so forth.

Note that closeness to blood supply is a leading requirement in our set of criteria, so that it is very likely that solution of the optimization problem on the basis of introduced criteria would provide higher value of the hepatocyte-sinusoid contact area, thus correcting lower (compared to experimental data) value of this parameter obtained in model [41]: "the representative model 2 showed a hepatocyte-sinusoid contact area of only $37.1 \pm 1.1\%$, which was significantly lower than the experimental situation $(48.5 \pm 2.5\%)$".

One of the advantages of the proposed optimization approach is that it allows incorporating requirements corresponding to different *scale levels*, such as the "smoothness" of the surface formed by hepatocytes in order to minimize the hydraulic resistance. Note that it is impossible to incorporate such a requirement into single-cell models, because it belongs to different (higher) dimensional scale level. Combining requirements and mechanisms corresponding to different scale levels (in other words, creating multiscale models) generally makes such models more adequate and comprehensive, and in many instances more manageable and robust, due to the possibility to introduce more constraints and interrelate model's parameters at different scale levels.

Another advantage of the proposed approach is that it can be easily combined with other models due to the multiscale nature of suggested method. It also allows simultaneously considering different types of cells (liver consists of several types of cells), which single-cell models are not adapted to do. Thus, a relatively simple consideration, which directly follows from the general growth law, allowed suggesting an efficient alternative approach for modeling hepatocytes growth with relation to specific features of vascular systems in livers.

Another interesting application of the general growth law that could make modeling of vascular systems more definitive and efficient is this. Since the general growth law allows finding metabolic characteristics of livers in terms of consumed nutrients, it potentially allows estimating the *overall* blood flow through the liver, if the amount of nutrients consumed by liver can be related to the overall blood flow. Given an important homeostatic role of the liver, such a possibility should not be ruled out, at least in certain scenarios.

Vascular system is important for VLN integral models as a supplier of nutrients and different substrates, such as drugs and other chemicals; it removes waste products; acts as a communication channel of an organism's signaling system, and supports other functions. It is through the blood flow that the limitations on the overall nutrient consumption are imposed, thus affecting metabolic rates, while concentration and composition of chemical substances in the blood largely defines specific features of biochemical reactions at different levels. The role of integral constraints and boundary conditions set by the blood flow, however obvious it may look, presently either is somewhat underestimated, or it is difficult to develop efficient methods to use the blood flow characteristics to the advantage of integral models. However, eventually, VLN model has to incorporate the blood flow and relate its characteristics to other model parameters. As the results of our study showed, the use of the general growth law for such purposes could be helpful.

# 3. Simulation Models

Most of these models, in one way or another, use stochastic approaches. Another classification distinguishes between individual based models (which can be lattice based or lattice free) and continuum models [42,43]. Review [13] considers many biomechanical models of growing tissues. It also introduces and analyzes methods based on cellular automata and off-lattice models. The idea behind cellular automata is this. The region of space is divided into lattice sites (one lattice can be associated with zero, one or several biological cells). During simulation, characteristics corresponding to lattices are tracked. For cell proliferation, such characteristic is if the lattice is occupied by a biological cell or not. Different enhancements of lattice models allowed accounting for cells' shape and size, certain mechanical effects [13]. However, as the authors of [13] say with regard to mechanical modeling, "none have satisfactorily dealt with all of the issues, and in particular the implementation of stress effects is a major difficulty and has not been achieved in a systematic or theoretically satisfactory way". The same is true with regard to many other growth characteristics, especially when it comes to growth aspects of the "second order", which, nonetheless, might provide growth effects of the "first order". For instance, in study [41], which uses off-lattice model with individual quasi-spherical cells, the authors discovered that "coordinated cell orientation and cell polarity were identified to be the most critical parameters. Elimination led to destruction of the characteristic micro-architecture of the lobule".

**3.1. Off-lattice models.** In the off-lattice models the characteristics of a cell ensemble are entirely determined by cells' position, which, in turn, are defined by interactions between cells. Cells are described by physical parameters derived from certain assumptions, which are not always well justified. Then, these parameters become entries to largely mechanistic models, which may account for cells' position, orientation, shape etc. Progress of off-lattice models is bound to more realistic input parameters and mathematical developments. In particular, introduction of the notion of interaction potential in [44] was beneficial both from mathematical perspective and also made models physically more meaningful.

Since off-lattice models presently receive attention, we will discuss them in detail. One of the important considerations that should be taken into account when considering these models for practical applications is that they are *very* computationally intensive. Authors of [43] emphasize this too saying "the simulation studies are highly computation-time consuming". The degree of computing intensity is such that even fast progress in computer technologies in the next decade probably won't be sufficient to satisfy the computational thirst of these models, unless some *qualitative* enhancements of models themselves are made.

It is probably tempting to use methods to computationally "regenerate" biological tissue, up to the whole organ, from few cells. Although on a smaller scale such models provide useful insights, such as the ones presented in [40–44], scaling such models to the organs' level is problematic for many reasons. Besides, there are alternative methods, which could allow obtaining similar results more efficiently. For instance, formation of "invasive fingers" at a *cell* scale, when cells proliferate on a plain surface [43], is an adequate application area for stochastic methods. On the other hand, formation of tissues is an area where the use of other methods seems more appropriate. First of all, tissues contain by orders of magnitude greater numbers of cells than present methods can handle. Secondly, tissues are composed of many types of cells of different sizes, whose organizational structure and mutual dependence and disposition is defined by many factors acting at *different* scale levels. This kind of information is difficult to incorporate into discussed models, while computational challenge will become insurmountable.

Integration of models with biochemical machinery, whose modeling is the primary goal of the VLN project (such as signaling,

gene regulatory and metabolic mechanisms) faces tremendous obstacles. The authors of analyzed models consider as an advantage that their models do not take into account nutrients, while explaining certain phenomena, like growth saturation, by purely mechanistic forces. However, nutrients is an extremely important factor in growth and functioning of all living organisms, and has to be incorporated into such comprehensive (from the biochemical perspective) models as VLN. In this regard, there are fairly advanced methods for modeling population dynamics, which take into account nutrient content and its distribution, as well as many other pertinent to population growth parameters. For instance, a general method proposed in [45]. Maybe adding such approaches could benefit the discussed simulation methods, since in this case the population quantity will be a known parameter derived from nutrients availability.

Adequately describing mechanical properties of cells is another serious problem. Results obtained in [41–44] suggest that compact cell aggregations experience rather strain and stress, which suppresses cell proliferation. This is true for tree trunks, whose internal annual layers experience stress [13], which is likely one of the main reasons why the wood inside the tree trunk is actually dead. However, in living organisms, unless this is pathology, cells are alive, which makes the assumption about internal tissue stress doubtful (and not only for this reason). Indeed, review [13] acknowledges that "it is stretch, rather than any other mechanical field (e.g., stress or strain) that the cells are "truly" experiencing - given that the proposed mechanisms all essentially depend on the deformation of protein molecules.".

**3.2. Area of application and limitations of off-lettuce models.** Mathematical base of individual-based off-lattice models is a non-parametric statistical model, which led the authors to use Metropolis algorithm. The standard Metropolis algorithm is a Markov chain Monte Carlo method to study equilibrium model with known Hamiltonian. Hamiltonian function characterizes *mechanical* systems through generalized coordinates, linear momentums and time. For instance, for a conservative system, Hamiltonian function is equal to a sum of kinetic and potential energy. It should always be kept in mind when using Hamiltonian for other than purely mechanical systems that Hamiltonian was introduced exclusively and makes physical sense only for *mechanical* systems, and so it can provide accurate description of system behavior *only* for them. Any other usage of Hamiltonian function is based on *analogy*, somehow found *similarity* of analyzed systems to purely mechanical ones. Cells, biological tissues definitely posses mechanical properties. However, these are biological mechanisms whose workings predominantly define mechanical properties of organisms and their constituents within the limits imposed by laws of mechanics, but not the other way around. Thus, it is logical that the core of biological models should be composed of biological mechanisms.

Metropolis algorithm was proposed for drawing samples from the physical distribution of thermal motion of molecules (Boltzmann distribution) for systems in *equilibrium*. Growing populations of cells cannot always be described this way. Given the fact that the cells' motion is determined by many factors, which dynamically change all the time, it is understandable that cells' motion is not really analogous to thermal motion of molecules, at least in the whole range of growth conditions.

The use of Metropolis algorithm implicitly assumes that the model parameters have a complex distribution. Otherwise, more simple methods for parameter estimation could be used. Therefore, a large quantity of uncertainty is present in the model from the outset. How this uncertainty relates to adequate modeling of cell proliferation is not clear and no analysis was done in this regard. As another line of inquiry, it would be interesting to consider a state space model for the cell proliferation process and use Markov chain Monte Carlo for inference in such a model [46].

Authors of review [13] express similar concerns about stochastic component of models: "Some of the off-lattice models studied are modeled using Monte Carlo simulations, which implicitly include a stochastic component for the cells' displacement. Is the inclusion of stochasticity relevant here? Experiments show that isolated cells are affected by Brownian motion... However, we would expect that in closely packed cell ensembles, the magnitude of the interaction forces would outweigh the stochastic component".

Extracellular matrix is an important factor in tissue growth, whose inclusion into adequate models of biological tissues is required, but such a task looks problematic for the discussed algorithms.

However, the most critical barrier is that such models *fundamentally* cannot naturally aspire to the scale of a whole organ, for several reasons. One is that they do not have mechanisms allowing the creation of structures. At the same time, living tissues and organisms in general are first of all about *structure*, and the structure which is diverse in content and organization, dynamic, complicated and whose constituents are enormously interdependent with interconnections at all scale levels.

This brief and far from being complete review shows that simulation methods can provide useful insights at a cellular level. However, scaling their area of applicability up is rather problematic. This inference is similar to what the authors of review [13] think: "However, accurate models of cell-cell interactions do have their place: on the cell scale".

# 4. Summary of Features of Existing Models

The analysis of available models which are supposed to contribute to VLN project and probably form its "skeleton" and "tissue" convincingly demonstrated the following.

1. Models capable of producing meaningful practical results related to the goal of VLN project (which is "to represent and simulate organ function" in order to address "the needs of patients and clinical practitioners alike"), have fewest abstraction layers, dealing directly (or with minimum intermediate layers) with values representing *real* physical, chemical, biological and other parameters.

2. The more common features models have, the more naturally their integration goes. This is especially true for features that form the foundation of models.

3. The better is experimental verifiability of models, the better models perform.

4. The more experimentally measured common or related parameters (both input and output) integrated models have, the better the resulting integral model is.

5. Simulation models based on stochastic approaches, such as individual-based off-lattice models, can provide useful insights on cellular level when cells begin to proliferate. However, their potential is somewhat overrated and the usage of such models at higher scale levels, at least at their present state, looks problematic, for which many objective reasons exist.

6. Models that have more deterministic features and are based on more deterministic algorithms perform considerably better, in all respects, than models based on complicated stochastic approaches. This is understandable, given the fundamentally deterministic nature of growth and after-growth existence of organs, systems and organisms.

Although there is certain randomness in working of biochemical, biophysical and others mechanisms which support life cycle of living organisms, these are effects of the second order, while biological mechanisms are fundamentally *deterministic*. If we think for a moment, it cannot be otherwise, since only a deterministic way of performing functions by cells, organs, systems can guarantee organisms' homeostasis and highly balanced and coordinated work of their numerous interconnected and highly dependable mechanisms and systems.

7. All models use such approaches, concepts and formal mathematical apparatuses that they could significantly benefit if more well defined constraints can be imposed. Especially noticeable is insufficient amount of *integral* constraints, which are required by many presently used modeling methods in order to "squeeze" solutions into acceptable error margins.

8. Present integration models unite particular models in a straightforward manner by linking them through their "outer interfaces". At the same time, more elaborate integration structure, such as hierarchical organization of models with different branches, model inclusion, etc. generally allows developing substantially more efficient integral models.

9. None of the biological mechanisms (such as biochemical or biophysical), on which considered models are based, have presence at all or at least several scale levels, although existence of such mechanisms would significantly facilitate development of integration models through the common foundation. We discussed the grounds for existence of such mechanisms and also found that the general growth law is a good candidate for such a unifying role across different scales.

10. Most models are based on biomolecular vision of biological phenomena; less frequently modeling is done at cells' level. Except for the models of vascular systems and a few other developments, the overall thinking paradigm is grounded into biomolecular understanding of life phenomena. This vision is in drastic contrast with inorganic matter, whose behavior is defined by so many *fundamental* laws of Nature acting at different spatial levels. It should not be the case that more complicated biological phenomena are governed predominantly by mechanisms at molecular level. There should be fundamental laws of Nature shaping living world at all higher than molecular levels.

## 5. Requirements and Criteria for Developing Biological Models

Using results of the presented analysis, we suggest that the integral biological model should satisfy to the following requirements.

1. Its core should be based on real biological mechanisms with as little abstraction layers as possible. (In this regard, we would like to quote the following: "The best conducted human activities are those which most faithfully resemble the operations of the natural world." -Wilhelm von Humboldt, "The limits of state action".)

2. The best approach to modeling living organisms would be on the basis of some real mechanisms (or better off, fundamental laws of Nature) that are common across different *scale* levels. Given the arrangement of natural laws existing at all spatial levels in the better studied inorganic world, such mechanisms, governing development of organic life, exist in Nature. The general growth law is one of such fundamental laws of Nature, whose use can benefit biological models, including integral ones.

3. Verifiability of an integral model and models it encompasses should have higher priority. This includes experimental observations, as well as other direct and indirect means of verification, up to the level of general philosophical principles of validation of scientific hypotheses [47]. Importance of experimental measurability follows from the general principle that practice is a criterion of truth. Models have to be *fundamentally* entirely verifiable. At the moment of creation, it can be a hypothetical model, but it should not include fundamentally unverifiable concepts, assumptions, etc., or parameters whose valuation is practically infeasible.

4. All other conditions being equal, priority should be given to models that include more directly measured parameters.

5. The models to be integrated should have as many common concepts and common or related parameters as possible.

6. Unless the inherent nature of some phenomenon is stochastic, priority has to be given to deterministic models, since biological phenomena are *fundamentally* deterministic. Although in many instances we may not know the cause of such determinism, it is there.

7. Models should have a sufficient amount of constraints, including *integral* ones, at all spatial levels, in order to "squeeze" solutions into acceptable for practical purposes error margins. Formal apparatuses (mathematical, logical, etc.) should have ability to incorporate additional information and constraints, which may appear later. Examples of such approaches can be found in [48].

8. Models, both integral and of lesser generality, should have optimal structure, including elaborate hierarchical organization of included models, with different branches, model inclusions, aggregations, etc. Such an approach generally allows developing more efficient integral and simpler models, than straightforward linking of different models through their outer interfaces.

9. Biological models should be based on a truly "multiscale" cognitive paradigm, which means that, generally, all scale levels should have equal importance and representation in models of biological processes, similarly to inorganic matter. This means that each scale level should have sufficiently developed specific set of concepts, methods and associated fundamental laws.

10. Based on experience in other disciplines, like physical sciences, software system development, etc. [28–33], as well as results of analysis of available integral models [34–37], "vertical" integration should be of higher priority. It should not be a complete model first, but rather a model's skeleton "vertically" piercing all scale levels. It should have scalable design that allows adding and interconnecting additional components and whole "branches" of more particular models.

Using these requirements as guidance, and results of analysis done in previous sections, the following development framework can be proposed for modeling biological phenomena, and liver in particular.

## 6. Framework for Modeling Biological Phenomena

The inference from the study above is that metabolism model should form the core vertical "stem" of integral biological models. There are several arguments that support this suggestion.

1. Metabolism is what makes living organisms truly "living". This is their most essential, of fundamental importance, property. No metabolism - no living creatures.

2. Existing metabolism models, including discussed above, are well developed. They can be scaled "vertically" using fluxes as a common base for different spatial levels. Metabolic models are scalable, which allows adding more information and details without jeopardizing their accuracy. Moreover, the more details such models incorporate, the more accurate they potentially become. This is an extremely important property, since most mathematical methods diverge when the number of parameters increases. Example can be Recon 2 project [36] for human metabolism, when more details allowed to significantly reduce the amount of dead-end metabolites compared to the previous Recon 1 study. Also, the authors acknowledge, "Recon 2 carried a nonzero flux for all tasks, compared with Recon 1, which achieved this functionality for only 83% of the tasks". This advancement is largely due to adding more metabolic networks to the study. In addition to these attractive features, the output parameters of metabolic models are real physical values which can be directly used in many practical applications. This is especially true with regard to VLN project, for which biochemical side of the integration model is of primary importance. Note that metabolic models also satisfy other criteria and requirements listed above. Although increase of model parameters often leads to decrease of models' accuracy, this is not some unavoidable property of such models, but it is due exclusively to the restrictiveness of particular mathematical methods [48]. It means that more efficient mathematical methods are needed, while fundamentally there are no limitations on improving model accuracy. Besides, the existing mathematical methods can often benefit from combination of several conceptual approaches. For instance, regularization algorithms often help to significantly enhance accuracy.

3. Several studies convincingly showed that it is possible to successfully integrate metabolic and other models. For instance, "horizontal" integration of metabolic, gene regulatory and signaling networks was done in [37], while example of "vertical integration" is in [34]. Of course, due to conceptual and implementation heterogeneity of different models such integration may present a challenge. However, if we take a closer look, we can see that such difficulties were due to rather non-metabolic models. For instance, in case of integration of signaling and metabolic networks this is the logical binary nature of signaling network that compromises this union [37]. In this case, it happens because metabolic models have the least abstraction layers and stay closer to objective reality than the model of signaling networks. Note that this and other examples of difficulties in integrating metabolic networks with other models are not of a fundamental nature. This is just an inadequate compatibility of particular formal apparatuses, not more than that. The problem is finding more appropriate formal approaches, mathematical or others, which seems as a feasible task.

4. In the requirements section, the need to use *real* mechanisms present in Nature as a common base across different scale levels was stated. Biochemical reactions is one of such common bases, affecting organisms' functions at all spatial levels, to which other mechanisms relate. In fact, apart from purely electromagnetic pulses traveling in neurons, the low level foundations of all biological mechanisms are biochemical in nature. Generation of electromagnetic impulses is due to biochemistry as well [1,49]. *Bioelectromagnetism* and *biochemistry* do not have a sharp boundary between them. On the other hand, chemical reactions are based on electromagnetic forces. So, these two foundations of living organisms, bioelectromagnetism and biochemical reactions, through which metabolic activity is realized, indeed, present two common organisms' bases across different scale levels.

5. The third common base in all living organisms is the *general growth law*. This is a fundamental law of Nature, which governs development of a living organism at different spatial levels through biochemical mechanisms. In particular, as we showed in this article, certain metabolic properties of organisms and its constituents can be found on the basis of the general growth law at all scale levels. These are much needed parameters, through which new important constraints in metabolic and other models can be introduced, such as, for instance, amount of nutrients required for biomass production and maintenance for different types of cells, organs, systems and the whole organisms. Note that presently there are no other methods of finding these parameters other than through the general growth law. This fact significantly reinforces our suggestion to consider metabolism models as a "vertical" core of integral biological models.

6. Metabolic properties of organisms and their constituents are directly or intermediately linked to many other properties, for instance size, shape, consumed nutrient influx, rate of growth, etc., as well as closely connected to other biological mechanisms, such as gene regulatory.

7. Metabolic models provide direct interface to the outer environment, for instance, through receptors, extracellular matrix, vascular systems, etc.

8. Metabolic models represent the first line of formalism describing real phenomena, using real physical and chemical values that can be objectively evaluated. In other words, so to say, metabolic models sit "right on top" of objective reality.

9. Parameters of metabolic models can be directly measured or objectively evaluated indirectly through measurements, which makes them transparent.

10. Closeness of metabolic models to objective reality, measurability of their parameters, scalability and vertical interdependence at different scale levels make metabolic models highly *verifiable*, in all their cross-sections, which is an extremely important property of any model pretending to objectivity and practical application.

Figure 6 shows a diagram of framework for developing biological models. The degree of association and aggregation of modules can vary. In some instances, more particular model could provide just a few independent parameters as its output to be used by the integral model, while in other cases integrated models may have close association or aggregation, with many common parameters. Not all possible models and relationships are shown. For instance, the general growth law can provide data for different models; input data can be used by diverse models as well.

With regard to the general growth law, it provides important and, in many instances, indispensable parameters. Its role becomes even more important with increase of models' complexity. For instance, this is the situation when one considers parasite-host (P-H) system, and host contains drugs targeting the parasite, such as in [50] in case of malaria. Let us denote the amount of produced biomass as PG for the parasite (denoted by letter 'P') growth, and accordingly HG for the host growth, while masses of substances participating in maintenance are accordingly PM and HM ('G' stands for growth, 'M' for "maintenance"). These four parts may interact in different combinations, for instance, PG - HM, PG - HG - HM, PM - HM - HG - PG, etc. Such possible, and actually observed, variety of combinations significantly complicates the use of methods of metabolic flux analysis, while finding masses of participating components through the growth equation could significantly facilitate finding solutions by imposing mass constraints, which are the most critical ones for this method. Note that such an approach is of general importance and can be equally applicable to other bacteria residing in different hosts.

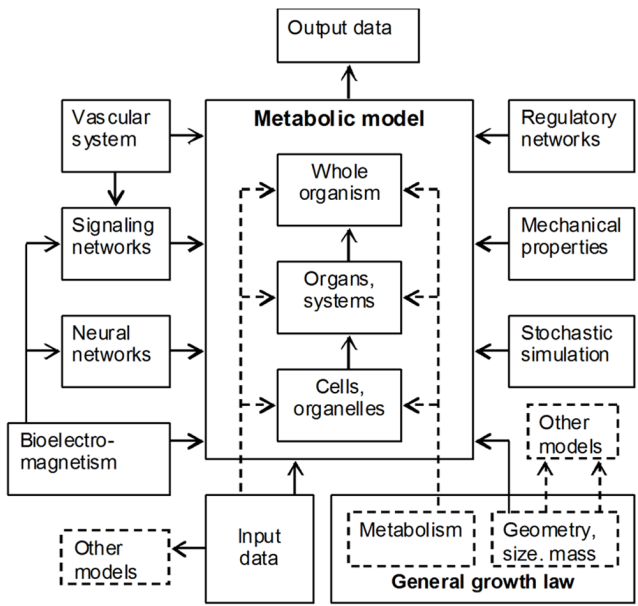

**Figure 6. Framework for developing biological models.**
doi:10.1371/journal.pone.0099836.g006

What about alternative core models? If we do a comparison based on introduced models and criteria, then we could see that metabolic model has a better rating than competing approaches. For instance, one of the major criteria applied to the development of integration models is availability of a common phenomenological base across as many included models as possible. In this regard, the metabolic model is preferred, since its foundation is formed by real biochemical mechanisms that are common across most models, and which underlie the working mechanisms at all spatial layers. Regulatory networks, one can say, also are widely present in different models and also represent real biochemical mechanisms. However, compared to the regulatory networks, the metabolic model has substantially more constraints which can be defined at all spatial levels (including the critical ones introduced through the general growth law). Besides, metabolic model has diverse relationships with other models and the surrounding environment. Overall, this makes metabolic models more robust, stable, scalable and accurate.

**6.1. Importance of including growth phase into integral models.** An important aspect of the proposed framework is its *dynamic* nature. It allows modeling the *whole* life cycle of all organism's constituents, for which the growth phase, as our study of liver metabolism showed, is of *paramount* importance. It is interesting that not much attention is given to modeling of growth phase in all discussed models. In our view, this is a major oversight, which has rational explanations, one of which is the difficulty of modeling dynamic processes in general and growth in particular. However, as we have seen, it is only through the modeling of growth phase that we were able to evaluate metabolic characteristics of a *fully grown* liver. If not for the growth equation and the growth ratio parameter in particular, we would not find nutrient consumption and other metabolic parameters at all. Knowing growth characteristics is as important for successful modeling as knowing how an individual grew up and in what environment, in order to make a judgment about the personality. It is very likely that a similar situation will be observed in modeling other organisms' constituents and whole organisms. It is no doubt that a correctly described growth process will provide lots of important insights into the nature of considered biological phenomena.

**6.2. Scale levels.** Scale levels also should reflect on real phenomena as closely as possible and have the fewest abstraction layers. There is no really well defined classification of scale levels in biology; various authors may mean different things. There is definitely biomolecular level. However, what size should biomolecular aggregations have in order to classify them as belonging to this scale too? The general criterion, which some theories of scientific knowledge provide, and dialectics in particular [47], is when accumulation of *quantitative* changes transforms into a new *quality*. In our case, that could mean appropriate introduction of new qualitative methods to deal with this new quality. For instance, when molecules assemble into ribosome, this aggregation acquires a new quality associated with a specific function performed by ribosome. Thus, in this case, organelles should be considered as belonging to the next scale level. Then cells' level follows, after which some functional tissue formations should be distinguished as the next scale level. Organs and systems usually unite several such tissue formations which together provide a qualitatively new function (or functions). Then the formation of several organs may be considered, although it seems like in many cases this would be a somewhat artificial creation, after which the level of whole organism follows.

**6.3. Liver model.** All considerations above are applied to liver modeling. Metabolism is especially suited as a "vertical" core for such a "porous" (with regard to fluid flow) organ as liver is. Vascular system is of high importance for liver function, as well as extracellular and intracellular flows. As we could see from the study of liver metabolism conducted in this paper on the basis of the general growth law and its mathematical representation the growth equation, it is possible to find integral metabolic characteristics. These characteristics are very important for developing metabolic models, whose accuracy and objectivity is as good as models' constraints are.

Gene regulatory networks apparently can be incorporated into the integral model on the metabolic basis. From the perspective of model design, signaling networks could be, to large extent, self-contained, but it depends on the level of details one needs. The impression is that the signaling model requires attention and probably qualitative breakthroughs which could lead to appearance of conceptually new approaches.

We won't go into further details, but the said demonstrates that the proposed framework is suitable for VLN project. Of course, as we stressed above, the model should be dynamical, and the growth phase should be an important and inherent part of it. As the previous studies and research presented in this paper showed, the general growth law could provide very important inputs into biological models in many aspects, from geometrical and volumetric characteristics and truly dynamical description of growth processes to finding metabolic characteristics. As a relatively new development, the wide application of the general growth law will require some efforts. The comprehension of the general growth law, and especially its conceptual understanding, is not as simple as it may seem on the surface. However, for the task of modeling, this is a *tool*, and, as every tool, it does not require the user to entirely understand how it is designed and what laws of nature it is based upon.

## Conclusion

From the standpoint of classification of scientific knowledge and considering the maturity of foundations underlying a scientific discipline, especially in the form of fundamental laws of Nature, currently biology as a science is on threshold of its next more mature phase. How long the process of entering into this new phase will take, strongly depends on development of general theories and high level generalizations, preferably in the form of fundamental laws similar to good times in physics in 20th century. Excitements of discoveries of new biomolecular mechanisms, especially the DNA's sprang studies, much stimulated by technological advances that opened gateways into a micro-world, eventually have to give way more rational attitude and allow development of other conceptually different approaches and paradigms, in particular at higher scale levels. Study of systemic level biological mechanisms is also such an area. Research diversification that allows to systematically and comprehensively study the scope of an entire discipline, if done correctly, will significantly accelerate development of biology as a truly scientific discipline. Besides, this will undoubtedly stimulate the process of convergence of numerous but fragmented biomolecular and others mechanisms, including those of higher spatial level, into a general biological framework.

In this work, we proposed method for finding metabolic characteristics of growing cells, organism's organs and systems on the basis of the general growth law, which is a fundamental law of Nature much responsible for the origin and evolutionary development of organic life and growth of individual organisms and their constituents at all scale levels. The proposed approach is also applicable to whole organisms. A validation method has been introduced in order to independently verify metabolic characteristics obtained by the method. We demonstrated application of the proposed method by finding metabolic characteristics of the dog and human livers and liver grafts and remnants. In particular, we found nutrient influx required to support maintenance needs and biomass production, as well as the total amount of nutrients required to support these activities for a certain period of time. We obtained the following important practical and theoretical results:

1. Using experimental data for two dogs, we found metabolic influxes per unit of volume for maintenance and biomass synthesis, depending on time. It was discovered that dogs' livers do have very high metabolic rates of $0.624 g/(cm^3 \times day)$ and $1.01 g/(cm^3 \times day)$.

2. Although we could not obtain accurate and reliable estimation of metabolic rates for human livers due to unavailability of data for individual liver remnants, we found that it is several times less than in dogs. For males, we obtained that the overall required nutrient influx is of the order of $0.088 g/(cm^3 \times day)$, and for females $0.098 g/(cm^3 \times day)$.

3. Obtained results indicate that females' liver metabolism is higher than males' (in terms of nutrient influx required to

support liver metabolism by about 10%). Higher metabolic activity of livers in females can be explained by evolutionarily developed reserve of liver metabolic capacity required to support child bearing. This discovered property of female livers is well supported by clinical observations. They show smaller size of grown female transplants and remnants, which is most likely explained by higher metabolic capacity of female livers when small grown livers can support normal liver function, so that there is no physiological need for bigger livers.

Although the study has been restricted to modeling growth of livers and finding liver metabolic properties, the area of application of the proposed method is by no means restricted to livers. The same approach is equally applicable to cells, cellular components, organisms' organs and systems, and whole organisms. In case of certain cells, such as fission yeast or *E. coli*, it is possible to determine more specific distribution of nutrients, that is influxes of nutrients that are used for protein synthesis, RNA synthesis and DNA synthesis at each phase of growth and during the overall growth period. Such a possibility is supported by the property of these organisms that different cell components have different rates of synthesis and degradation, which allows distinguishing nutrient consumption between components. This topic was considered in the presently submitted article.

Overall, the presented study of metabolic properties of livers on the basis of the general growth law and its mathematical representation the growth equation convincingly proved that biophysical mechanisms acting at higher than molecular levels provide valuable and in some instances indispensable input to biological studies. Many presently undertaken projects, aiming at practical biological and medical applications at organ, system or organism's level, faced serious obstacles trying to solve their problems using approaches based on low level biochemical and cellular mechanisms. Although some researchers think that these are just technical issues, in fact, the origin of these problems is in underdeveloped fundamentals and somewhat aberrated general scientific methodologies. In this regard, we suggest to consider the study as a pilot project opening the gates from the largely biomolecular playground into the entire world of multiscale biological mechanisms and fundamental laws.

The study also enhanced understanding of the general growth law as a fundamental law of nature that governs evolutionary development and growth and replication of individual organisms and its constituents. Previously, the validity of this law was convincingly demonstrated by application to unicellular organisms. The presented results of application of the general growth law to a problem of modeling growth of *organs*, in particular transplanted livers in dogs and humans, one more time convincingly proved both validity of the general growth law and its high practical and theoretical value for finding metabolic characteristics of living organisms and their constituents at organ level, which is a significant advance in development of the general theory of growth and replication of living organisms.

We conducted detailed analysis of available biological models and approaches from the perspective of creating integral models of biological phenomena. Based on this analysis and results obtained in this paper, we first introduced a generalized set of requirements and criteria which biological models should satisfy, and then introduced a general framework for development of biological models and provided supporting argumentation. The core of the framework, in our opinion, should be a vertical (meaning scale levels) metabolic model which integrates particular models. Brief analysis showed that the suggested modeling framework can be used as a base for VLN project.

## Acknowledgments

Author thanks A. Y. Shestopaloff for discussions, comments and editing efforts, Dr. P. H. Pawlowski from Institute of Biochemistry and Biophysics, PAS, for enduring support of the study of the general growth law and enforcing an idea to relate the general growth law to metabolic characteristics of organisms in general, Dr. I. F. Sbalzarini from Max Planck Institute of Molecular Cell Biology and Genetics (CBG) for discussions and project support, Librarian S. Thüm from CBG for the help in finding materials, Editor of ''Biophysical Journal'' Dr. P. Hunter for the recommendation to relate the study to Virtual Liver Network project, Editor Dr. M. Vinciguerra for the fast and efficient editorial process, and Reviewers for valuable comments.

## Author Contributions

Analyzed the data: YS. Contributed reagents/materials/analysis tools: YS. Contributed to the writing of the manuscript: YS. Developed software for analysis and modeling: YS.

## References

1. Shestopaloff YK (2014) Growth as a union of form and biochemistry. How the unity of geometry and chemistry creates living worlds through fundamental law of nature - the general growth law. Fourth revised edition. AKVY Press, Toronto. 455 p.
2. Kam I, Lynch S, Svanas G, Todo S, Polimeno L, et al. (1987) Evidence that host size determines liver size: studies in dogs receiving orthotopic liver transplants. Hepatology 7(2): 362–366.
3. Pomfret EA, Pomposelli JJ, Gordon FD, Erbay N, Price LL, et al. (2003) Liver regeneration and surgical outcome in donors of right-lobe liver grafts. Transplantation 76(1): 5–10.
4. Haga JM, Shimazu G, Wakabayashi M, Tanabe M, Kawachi S, et al. (2008) Liver regeneration in donors and adult recipients after living donor liver transplantation. Liver Transplantation 14: 1718–1724.
5. Holzhütter HG, Drasdo D, Preusser T, Lippert J, Henney AM (2012) The virtual liver: a multidisciplinary, multilevel challenge for systems biology. Wiley Interdiscip Rev Syst Biol Med 4(3): 221–235.
6. Berg JM, Tymoczko JL, Stryer L (2006) Biochemistry. 6th ed. W. H. Freeman, New York. 1120 p.
7. Mitchell S, Mendes P (2013) A Computational model of liver iron metabolism. arXiv: 1308.5826 [q-bio.MN].
8. Calvetti D, Kuceyeski A, Somersalo E (2008) A mathematical model of liver metabolism: from steady state to dynamic. Journal of Physics: Conference Series. 241(1).
9. Matthias K, Holzhuetter H-G (2012) Kinetic modeling of human hepatic glucose metabolism in T2DM predicts higher risk of hypoglycemic events in rigorous insulin therapy. J Biol Chem 287: 36978–36989.
10. Guillouzoa A, Corlub A, Aninata C, Glaise D, Morel F, et al. (2007) The human hepatoma HepaRG cells: A highly differentiated model for studies of liver metabolism and toxicity of xenobiotics. Chemico-Biological Interactions 168(1): 66–73.
11. Furchtgott LA, Chow CC, Periwal V (2009) A model of liver regeneration. Biophys J 96: 3926–3935.
12. Shestopaloff YK (2012) Predicting growth and finding biomass production using the general growth mechanism. Biophysical Reviews and Letters 7(3–4): 177–195.
13. Jones GW, Chapman SJ (2012) Modeling Growth in Biological Materials. SIAM Review 54(1): 52–118.
14. Drasdo D, Hoehme S (2012) Modeling the impact of granular embedding media, and pulling versus pushing cells on growing cell clones. New Journal of Physics 14: 055025.
15. Shestopaloff YK (2012) General law of growth and replication, growth equation and its applications. Biophysical Reviews and Letters 7(1–2): 71–120.
16. Shestopaloff YK (2012) The law of replication and growth. Almanac ''Lebed''. No. 665. Lebed website. Available: http://www.lebed.com/2012/art6096.htm. Accessed 2014 May 21.
17. Shestopaloff YK (2010) The role of physical and geometrical factors in the growth of living organisms. Biophysical Reviews and Letters 5(1): 43–58.
18. Shestopaloff YK (2011) A mathematical model of the physical growth mechanism and geometrical characterization of growing forms. International Journal of Biomathematics 4(1): 35–53.
19. Shestopaloff YK (2010) Physics of growth and replication. Physical and geometrical perspectives on living organisms' development. AKVY Press, Toronto. 174 p.
20. Tyson JJ, Csikasz-Nagy A, Novak B (2002) The dynamics of cell cycle regulation. BioEssays 24: 1095–1109.
21. Stephanopoulos GN, Aristos AA, Nielsen J (1998) Metabolic engineering: principles and methodologies. Academic Press, New York. 725 p.
22. Mahadevan R, Edwards JS, Doyle III FJ (2002) Dynamic Flux Balance Analysis Approaches. Biophys J 83: 1331–1340.
23. Maaloe O, Kjeldgaard NO (1966) Control of macromolecular synthesis; a study of DNA, RNA, and protein synthesis in bacteria. W. A. Benjamin, New York. 284 p.
24. Mente C, Prade I, Brusch L, Breierb G, Deutsch A (2012) A lattice-gas cellular automaton model for in vitro sprouting angiogenesis. Acta Phys Pol B 5(1): 99–115.
25. (2008) The liver: Biology and pathobiology. Ed. by Arias I M, Wolkoff A, Boyer J, Shafritz D. John Wiley & Sons, Ltd, Chichester, UK. 1216 p.
26. Pollitzer E (2013) Cell sex matters. Nature (1 August) 500: 23–24.
27. Shestopaloff YK (2010) Properties and interrelationships of polynomial, exponential, logarithmic and power functions with applications to modeling natural phenomena. AKVY Press, Coral Springs. 230 p.
28. Shestopaloff YK (2011) Polarization invariants and retrieval of surface parameters using polarization measurements in remote sensing applications. App Optics 50(36): 6606–6616.
29. Shestopaloff YK (2011) Distributed parametric effect in long linas and its applications, Int J Electronics 98(10): 1433–1443.
30. Shestopaloff YK (1993) Statistical processing of passive microwave data. IEEE Trans on Geosci and Remote Sensing 31(5): 1060–1065.
31. Shestopaloff YK (2011) Properties of sums of some elementary functions and their application to computational and modeling problems. J Comp Math and Math Physics 51(5): 699–712.
32. Shestopaloff YK (2012) Conceptual framework for developing and verification of attribution models. Arithmetic attribution models. The Journal of Performance Measurement 17(1): 48–59.
33. Shestopaloff YK (2011) Design and implementation of reliable and high performance software systems including distributed and parallel computing and interprocess communication designs. AKVY Press, Coral Springs. 226 p.
34. Markus K, Schaller S, Borchers S, Glaise D, Morel F, et al. (2012) Integrating cellular metabolism into a multiscale whole-body model. PLoS Comput Biol 8(10): e1002750. doi:10.1371/journal.pcbi.1002750.
35. Díaz Ochoa JG, Bucher J, Péry AR, Zaldivar Comenges JM, Niklas J, et al. (2013) A multi-scale modeling framework for individualized, spatiotemporal prediction of drug effects and toxicological risk. Front Pharmacol 3: 204. doi:10.3389/fphar.2012.00204.
36. Thiele I, Swainston N, Fleming RM, Hoppe A, Sahoo S, et al. (2013) A community-driven global reconstruction of human metabolism. Nat Biotechnol 31(5): 419–25.
37. Gonçalves E, Bucher J, Ryll A, Niklas J, Mauch K, et al. (2013) Bridging the layers: towards integration of signal transduction, regulation and metabolism into a mathematical model. Mol BioSyst 9: 1576–1583.
38. Grabin VG (1989) Oruzhie pobedy (Weapon of victory). Izdatel'stvo politicheskoi literatury, Moscow, Russian edition. 260 p.
39. Schwen LO, Preusser T (2012) Analysis and Algorithmic Generation of Hepatic Vascular Systems. International Journal of Hepatology 2012: 1–17.
40. Hoehme S, Brulport M, Bauer A, Bedawy E, Schormann W, et al. (2010) Prediction and validation of cell alignment along microvessels as order principle to restore tissue architecture in liver regeneration. Natl Acad Sci U S A 107(23): 10371–10376.
41. Byrne H, Drasdo D (2009) Individual-based and continuum models of growing cell populations: a comparison. J Math Biol 58(4–5): 657–87.
42. Drasdo D, Hoehme S (2012) Modeling the impact of granular embedding media, and pulling versus pushing cells on growing cell clones. New Journal of Physics 14: 055025.
43. Höhme S, Hengstler JG, Brulport M, Schafer M, Bauer A, et al. (2007) Mathematical modelling of liver regeneration after intoxication with CCl(4). Chem Biol Interact May 20; 168(1): 74–93.
44. Drasdo D, Hohme S (2005) A single-cell-based model of tumor growth in vitro: Monolayers and spheroids. Phys Biol 2: 133–147.
45. Shestopaloff YK (2013) A general method for modeling population dynamics and its applications. Acta Biotheoretica 61(4): 499–519.
46. Shestopaloff AY, Neal RM (2013) MCMC for non-linear state space models using ensembles of latent sequences. University of Toronto website. Available: http://www.utstat.toronto.edu/~alexander/Accessed 2014 May 21.
47. Shestopaloff YK (2011) Hypotheses validation by dialectical laws. Shestopaloff website. Avaialble: http://www.shestopaloff.ca/yuri_eng/natural_philosophy/003InquiryHypothesesValidation4.pdf Accessed 2014 May 21.
48. Bogorodsky VV, Kozlov AI, Shestopaloff YK (1985) Two approaches to object identification using microwave radiometry. Soviet Physics. Technical Physics 30(10): 1236–1237.
49. Malmivuo J, Plonsey R (1995) Bioelectromagnetism. Principles and applications of bioelectric and biomagnetic fields. Oxford University Press, Oxford. 482 p.
50. Huthmacher C, Hoppe A, Bulik S, Holzhütter H-G (2010) Antimalarial drug targets in Plasmodium falciparum predicted by stage-specific metabolic network analysis. BMC Syst Biol 4: 120.